\pdfoutput=1

\documentclass[11pt]{article}

\usepackage[preprint]{acl}

\usepackage{times}
\usepackage{latexsym}

\usepackage[T1]{fontenc}

\usepackage[utf8]{inputenc}

\usepackage{microtype}

\usepackage{inconsolata}

\usepackage{graphicx}

\usepackage{amsmath,amsfonts,bm}

\def\eqref#1{equation~\ref{#1}}

\def\1{\bm{1}}

\DeclareMathAlphabet{\mathsfit}{\encodingdefault}{\sfdefault}{m}{sl}
\SetMathAlphabet{\mathsfit}{bold}{\encodingdefault}{\sfdefault}{bx}{n}

\usepackage{hyperref}
\usepackage{url}

\usepackage{xspace}
\usepackage{amsmath}
\usepackage{graphicx}
\usepackage{hyperref}
\usepackage{booktabs}
\usepackage{adjustbox}
\usepackage{multirow}
\usepackage{tabularx}
\usepackage{booktabs}
\usepackage{caption}  
\usepackage[inkscapelatex=false]{svg}
\usepackage{subfigure}
\usepackage{subcaption}
\usepackage{listings}

\newcommand{\llm}{LLM\xspace}
\newcommand{\codex}{\textsc{Codex}\xspace}
\newcommand{\codellama}{CodeLlama\xspace}

\newcommand{\deepseek}{DeepSeek-Coder\xspace}
\newcommand{\deepseekinstruct}{DeepSeek-Coder-Inst\xspace}
\newcommand{\codellamainstruct}{CodeLlama-Inst\xspace}
\newcommand{\wizardcoder}{{WizardCoder}\xspace}
\newcommand{\magicoder}{Magicoder\xspace}
\newcommand{\sweagent}{SWE-agent\xspace}
\newcommand{\autocoderover}{AutoCodeRover\xspace}

\newcommand{\humaneval}{\textsc{HumanEval}\xspace}

\newcommand{\mbpp}{{MBPP}\xspace}

\newcommand{\swebench}{SWE-bench\xspace}
\newcommand{\swebenchlite}{SWE-bench Lite\xspace}

\newcommand{\parabf}[1]{\vspace{1mm}\noindent\textbf{#1} \hspace{0.5em}}

\newcommand{\model}{CSR-Bench\xspace}
\newcommand{\agent}{CSR-Agents\xspace}

\usepackage{xcolor}      
\usepackage{tcolorbox}   
\usepackage{graphicx}    
\usepackage{caption}     

\definecolor{myboxcolor}{RGB}{240, 240, 255}   
\definecolor{myframe}{RGB}{0, 0, 128}          
\definecolor{my_green}{RGB}{34, 139, 34}       
\definecolor{my_yellow}{RGB}{255, 215, 0}      

\newtcolorbox{mybody}{
  colback=myboxcolor,
  colframe=myframe,
  boxrule=1pt, 
  left=1pt,
  right=1pt,
  top=1pt,
  bottom=1pt,
}

\newcommand{\green}[1]{\textcolor{my_green}{#1}}

\title{\raisebox{-12pt}{\includegraphics[height=38pt]{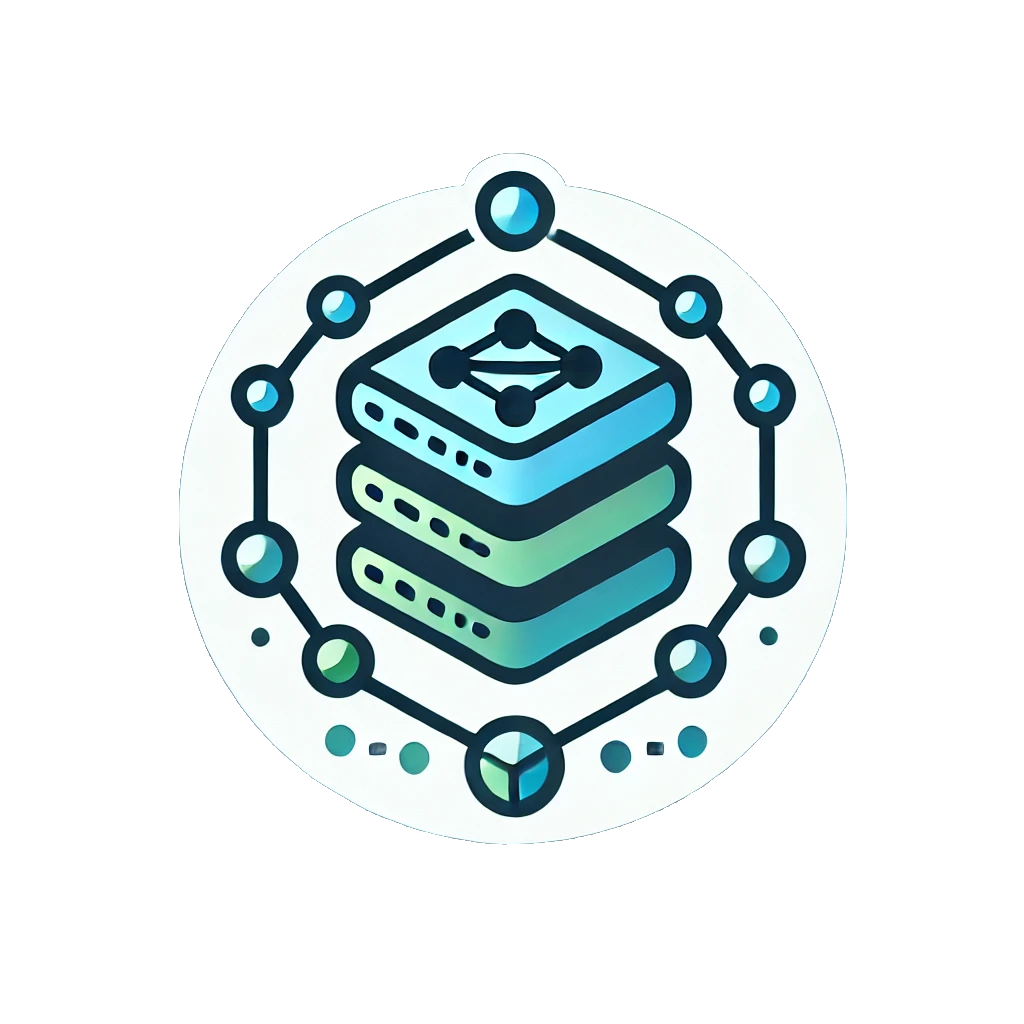}} \hspace{-0.2em} CSR-Bench: Benchmarking LLM Agents in Deployment of Computer Science Research Repositories}

\author{Yijia Xiao$^\dagger$*  \quad Runhui Wang$^\ddagger$*  \quad  \textbf{Luyang Kong}$^{\ddagger}$*  \quad 
\textbf{Davor Golac}$^\ddagger$ \quad \textbf{Wei Wang}$^\dagger$ 
  \\ 
$^\dagger$University of California, Los Angeles\\
$^\ddagger$ Amazon Web Services\\
{\small * Equal Contribution} \\
\texttt{\small \{yijia.xiao,weiwang\}@cs.ucla.edu} \\
\texttt{\small \{runhuiw,luyankon,dgolac\}@amazon.com}}

\begin{document}
{\makeatletter\acl@finalcopytrue
  \maketitle
}

\begin{abstract}
The increasing complexity of computer science research projects demands more effective tools for deploying code repositories. Large Language Models (LLMs), such as Anthropic Claude and Meta Llama, have demonstrated significant advancements across various fields of computer science research, including the automation of diverse software engineering tasks. 
To evaluate the effectiveness of LLMs in handling complex code development tasks of research projects, particularly for NLP/CV/AI/ML/DM topics, we introduce CSR-Bench, a benchmark for Computer Science Research projects. This benchmark assesses LLMs from various aspects including accuracy, efficiency, and deployment script quality, aiming to explore their potential in conducting computer science research autonomously. We also introduce a novel framework, CSR-Agents, that utilizes multiple LLM agents to automate the deployment of GitHub code repositories of computer science research projects. Specifically, by checking instructions from markdown files and interpreting repository structures, the model generates and iteratively improves bash commands that set up the experimental environments and deploy the code to conduct research tasks. Preliminary results from CSR-Bench indicate that LLM agents can significantly enhance the workflow of repository deployment, thereby boosting developer productivity and improving the management of developmental workflows. 
\end{abstract}

\section{Introduction} \label{intro}

With the rapid evolution of Large Language Models (LLMs), it has been demonstrated that LLMs have increased reasoning ability over the last few years, making intelligent agents based on LLM possible. Current agent-related applications in computer science include code writing (Codex~\cite{codex}, Deepseek Code~\cite{deepseek}, CodeLLaMA~\cite{codellama}, etc.), code base generation (MetaGPT~\cite{hong2023metagpt}, Agentless~\cite{xia2024agentless} and CodeStar\cite{starcoder}), code correction (SWEBench~\cite{swebench}), and more.

In computer science research projects, the associated codebases grow very rapidly, and a self-consistent codebase typically has several parts, including the instruction file (e.g., the README file), associated code packages, and related data. The instruction file usually contains an overview of the project, including environment setup, data preparation (e.g., data download and pre-processing, model weights download and preparation), model training process, performance evaluation, and the setup of a demo project (e.g., a chatbot project that interacts with users). In computer science, prestigious conferences, including NAACL, ACL, ICLR,  NeurIPS, CVPR, KDD, etc., encourage researchers to release source code for reproducibility of their accepted papers, and GitHub is the top choice for maintaining codebases for most researchers. A major step of computer science research is reproducing existing work, which is essential to gain insights and propose novel methodologies. However, even for well-documented and self-consistent projects, the setup process requires manual efforts and cannot be fully automated; many steps of setting up a code repository are rather mechanical, such as installing/updating dependency packages to configure the environment, downloading data, updating the relevant script/data directories, etc., which is tedious and often time-consuming. 

To tackle such challenges, we propose to use LLM agents for automating the deployment of code repositories of research projects, and build a benchmark, \textbf{C}omputer \textbf{S}cience \textbf{R}esearch \textbf{Bench}mark (\model) for evaluating LLM Agents on code repository deployment tasks. We also propose a multi-agent collaborative framework, \agent, to automate the deployment of code repositories by coordinating multiple LLM agents with different expertise and iterative improvement with provided tools. From more than ten major NLP/CV/AI/ML/DM conferences, we collected the 100 highly rated repositories\footnote{With appropriately permissive licenses.}, which are carefully selected from an initial candidate set that has more than 1500 top-star repositories. 
Our selection criteria include topic diversity and self-containment\footnote{The information within the repository is mostly sufficient for a successful deployment.} so that \model can provide a comprehensive evaluation of LLM agents on code deployment tasks including instruction generation, command execution, and self-improvement with tools.

To the best of our knowledge, \model is the first benchmark for the deployment of computer science research projects, providing a reference for evaluating LLM agents. We note that the success of code deployment depends not only on code generation, but also on many non-coding factors including tasks like experimental environment setup, data/model preparation, correcting bash commands, searching for solutions, and etc. Our proposed multi-agent framework, \agent, aims to achieve automation of code deployment, which can accelerate the progress for computer science research projects.

Our contributions are as follows:
\begin{itemize}
    \item We introduce \model to assess LLM's ability to understand instruction manuals and complex project structures, generate executable commands for code deployment, and solve errors during deployment.
    \item We propose the \agent framework, which leverages multi-agent cooperation with specialized capabilities including instruction comprehension, command execution, error log analysis, and error correction with searching and retrieval tools.
    \item We design a standardized testing system for reproducibility in \model, which can make \agent a CI/CD\footnote{Continuous Integration and Continuous Delivery.} standard system of computer science code repository deployment, ensuring ease of use, reusability, and improving communication and collaboration efficiency in computer science research projects.
    \item We evaluated a wide range of foundation models for \agent on \model. Results indicate that LLM agents can potentially accelerate the process of repository deployment, thereby boosting researcher productivity. However, it is still challenging to achieve full automation.
\end{itemize}

\section{Related Work} \label{related}







\parabf{Coding \llm{s}.}
Large Language Models (\llm{s}) have become the go-to solution for a wide array of coding tasks due to their exceptional performance in both code generation and comprehension~\cite{codex}. These models have been successfully applied to various software engineering activities, including program synthesis~\cite{patton2024programming, codex, li2022competition, iyer2018mapping}, code translation~\cite{pan2024lost, roziere2020unsupervised, roziere2021leveraging}, program repair~\cite{xia2023repairstudy, chatrepair, monperrus2018living, bouzenia2024repairagent}, and test generation~\cite{titanfuzz, fuzz4all, deng2023fuzzgpt, lemieux2023codamosa, kang2023testing}. Beyond general-purpose \llm{s}, specialized models have been developed by further training on extensive datasets of open-source code snippets. Notable examples of these code-specific \llm{s} include \codex~\cite{codex}, \codellama~\cite{codellama}, StarCoder~\cite{starcoder,starcodertwo}, and \deepseek~\cite{deepseek}. Additionally, instruction-following code models have emerged, refined through instruction-tuning techniques. These include models such as \codellamainstruct~\cite{codellama}, \deepseekinstruct~\cite{deepseek}, \wizardcoder~\cite{wizardcoder}, \magicoder~\cite{magicoder}, and OpenCodeInterpreter~\cite{zheng2024opencodeinterpreter}.

\parabf{Benchmarking \llm-based coding tasks.}
To assess the capabilities of \llm{s} in coding, a variety of benchmarks have been proposed. Among the most widely utilized are \humaneval~\cite{codex} and \mbpp~\cite{austin2021program}, which are handcrafted benchmarks for code generation that include test cases to validate the correctness of \llm outputs. Other benchmarks have been developed to offer more rigorous tests~\cite{evalplus}, cover additional programming languages~\cite{zheng2023codegeex,cassano2023multipl}, and address different programming domains~\cite{livecodebench, hendrycksapps2021, codecontest, ds1000, arcade}.

More recently, research has shifted towards evaluating \llm{s} on real-world software engineering challenges by operating on entire code repositories rather than isolated coding problems~\cite{swebench, zhang2023repocoder, liu2023repobench}. A notable benchmark in this area is \swebench~\cite{swebench}, which includes tasks requiring repository modifications to resolve actual GitHub issues. The authors of \swebench have also released a more focused subset, \swebenchlite~\cite{swebenchlite}, which contains 300 problems centered on bug fixing that only involves single-file modifications in the ground truth patches. ML-Bench \cite{liu2023mlbench} is a benchmark for evaluating large language models and agents for Machine Learning tasks on reporitory-level code. It involves 18 repositories and focuses on code generation and interactions with Jupyter Notebooks.

\parabf{Repository-level coding.}
The rise of agent-based frameworks~\cite{xi2023rise} has spurred the development of agent-based approaches to software engineering tasks. Devin~\cite{devinwebpage} (and its open-source counterpart OpenDevin~\cite{opendevin}) is among the first comprehensive \llm agent-based frameworks. Devin employs agents to first perform task planning based on user requirements, then allows them to use tools like file editors, terminals, and web search engines to iteratively execute the tasks. \sweagent~\cite{sweagent} introduces a custom agent-computer interface (ACI), enabling the \llm agent to interact with the repository environment through actions like reading and editing files or running bash commands. Another agent-based approach, \autocoderover~\cite{autocoderover}, equips the \llm agent with specific APIs (e.g., searching for methods within certain classes) to effectively identify the necessary modifications for issue resolution. Beside these examples, a variety of other agent-based approaches have been developed in both open-source~\cite{aidar} and commercial products~\cite{bouzenia2024repairagent, coder, repounderstander, lingma, factorydroid, ibmagent, opencsgstarship, marscode, amazonqdeveloper}.


Unlike existing benchmarks and agent-based frameworks, which focus on the code generation/completion tasks, our proposed \model and \agent focus on the code deployment task, which is under-studied in the field.
\section{\model} \label{gsrbench}
In this section, we provide the problem statement for code deployment in \model, introduce the code repository collection process of computer science research projects, and show their statistics.
\subsection{Problem Statement}
The \model consists of a collection of computer science research repositories from GitHub and these repositories are used for evaluating the capabilities of LLMs in code deployment tasks. For each repository, the deployment tasks typically include: (1) setting up the environment; (2) preparing necessary data and model files; (3) conducting model training; (4) demonstration of inference; (5) performance evaluation. To complete these tasks, we prompt LLMs to generate executable bash commands by using the README file as the primary source of information and other repository contents (source code, bash scripts, directory structure, and etc.) as supplementary information.




\textbf{Metric.}
During the evaluation in \model, the large language model will be prompted to generate executable commands for the corresponding sections for each repository. 
we use the completion rate as a key metric, defined as the ratio between number of successfully executed commands 
and the total number of commands executed.
\subsection{Repository Collection}
In \model, we aim to collect a diverse and comprehensive collection of code repositories of computer science-related research projects. GitHub is a good data source for this purpose and it provides tags for identifying most relevant repositories. Some example tags are “nlp”, “naacl”, and “emnlp2024”. Since \model focuses on computer science-related repositories, we filter the repositories by tags of various conference names and categories to ensure they include diverse topics. 

For repository selection, we use GitHub tags to obtain an initial set of over 1500 repositories that are relevant to computer science research topics and categorizing them into five areas: Natural Language Processing, Computer Vision, Large Language Models, Machine Learning, and Interdisciplinary topics. Notably, we collect repositories related to large language models because nowadays LLM-related research projects are increasingly popular due to its foundational impact in various areas of computer science. 

\begin{figure*}[htbp]
    \centering
    \begin{minipage}[b]{0.45\linewidth}
        \centering
        \includegraphics[width=\linewidth]{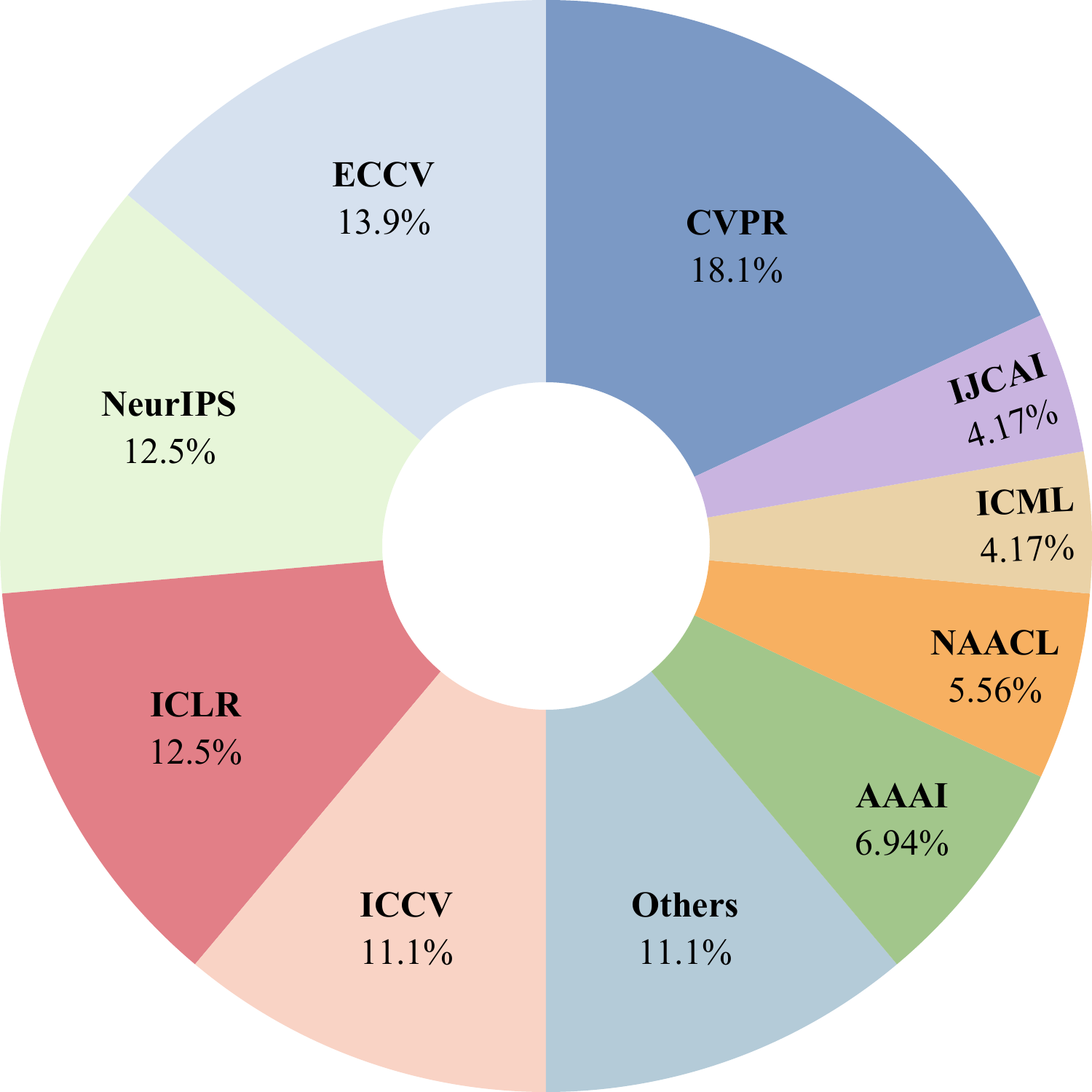}
        \caption{Conf Distribution of \model}
        \label{fig:conf}
    \end{minipage}
    \hfill
    \begin{minipage}[b]{0.45\linewidth}
        \centering
        \includegraphics[width=\linewidth]{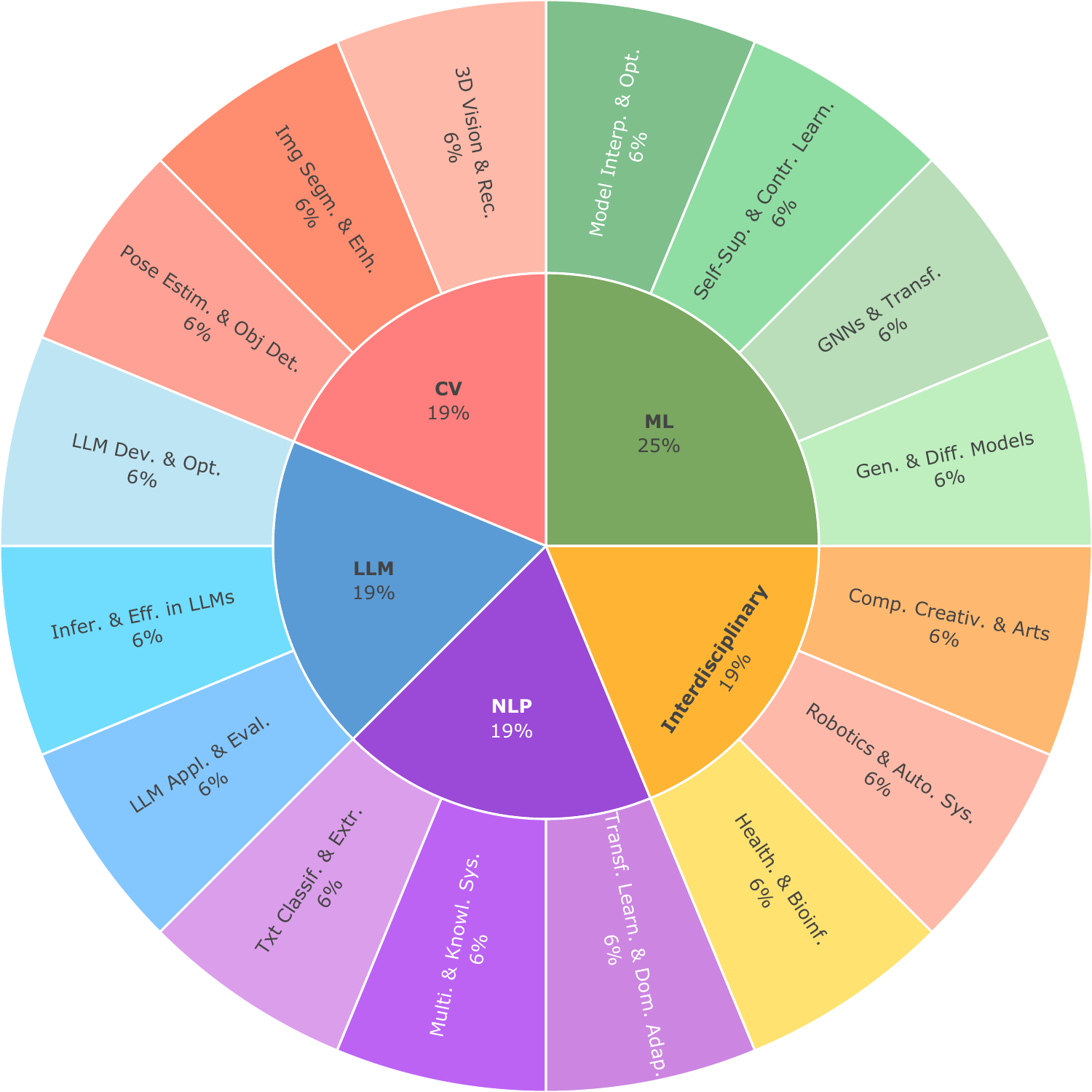}
        \caption{Topic Distribution of \model}
        \label{fig:topic}
    \end{minipage}
    \vspace{-1em}
\end{figure*}

We obtain 100 high-quality code repositories for \model in the following steps. First, we categorize this initial set and sort them by the number of GitHub stars. Next, we manually check the content of each repository starting from the top of the sorted list. In this step, we only keep repositories that contain sufficient information in their README files. We also skip the repositories that do not contain deployable code. Finally, we check the licenses of the repositories and make sure they are permissive.


\subsection{Statistics of \model}

This section provides an in-depth analysis of the traits of repositories in \model. We examine the diversity and breadth of topics covered, as well as detailed statistics about the documentation and structure of these repositories.

In \model, the README files and directory structures provide critical insights into the usability and organization of repositories. We use the following figures to analyze the lengths of README file and number of files, and offer a quantitative view of content complexity and organizational depth. The length of README file is an important metric because the most LLMs have limits on the input token length. The number of files indicate the complexity of the code repository.

\begin{figure}[ht]
    \centering
    \includegraphics[width=\linewidth]{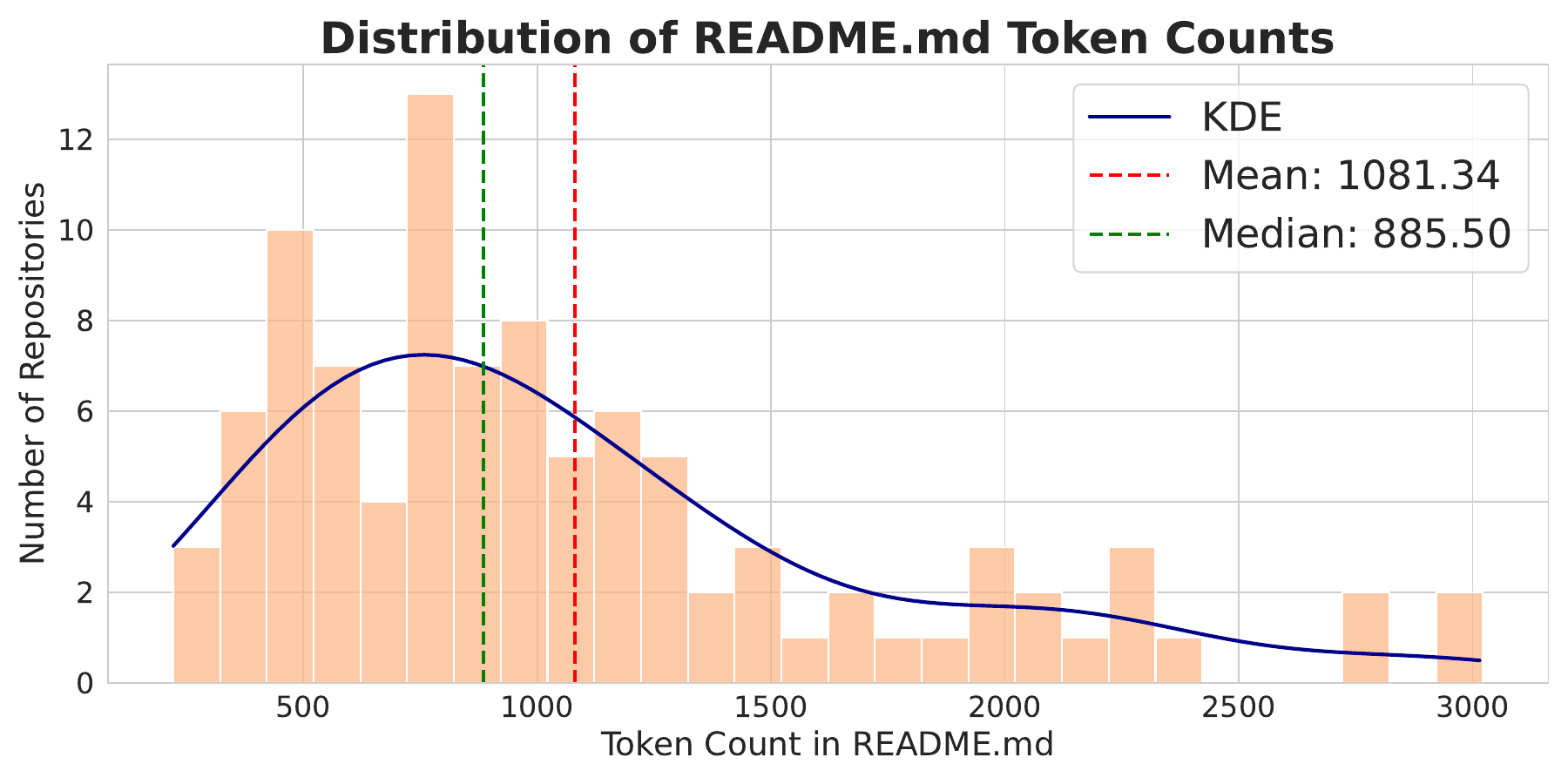}
    \vspace{-1.5em}
    \caption{Number of Tokens per README}
    \label{fig:token_count_distribution}
\end{figure}

Figure \ref{fig:token_count_distribution} shows the distribution of token counts in README files, highlighting the extent of documentation, which is essential for user understanding and repository usability. Since the mean token counts is just over 1000 and the maximum counts is around 3000, most LLMs can take the full README files as input.

\begin{figure}[ht]
    \centering
    \includegraphics[width=\linewidth]{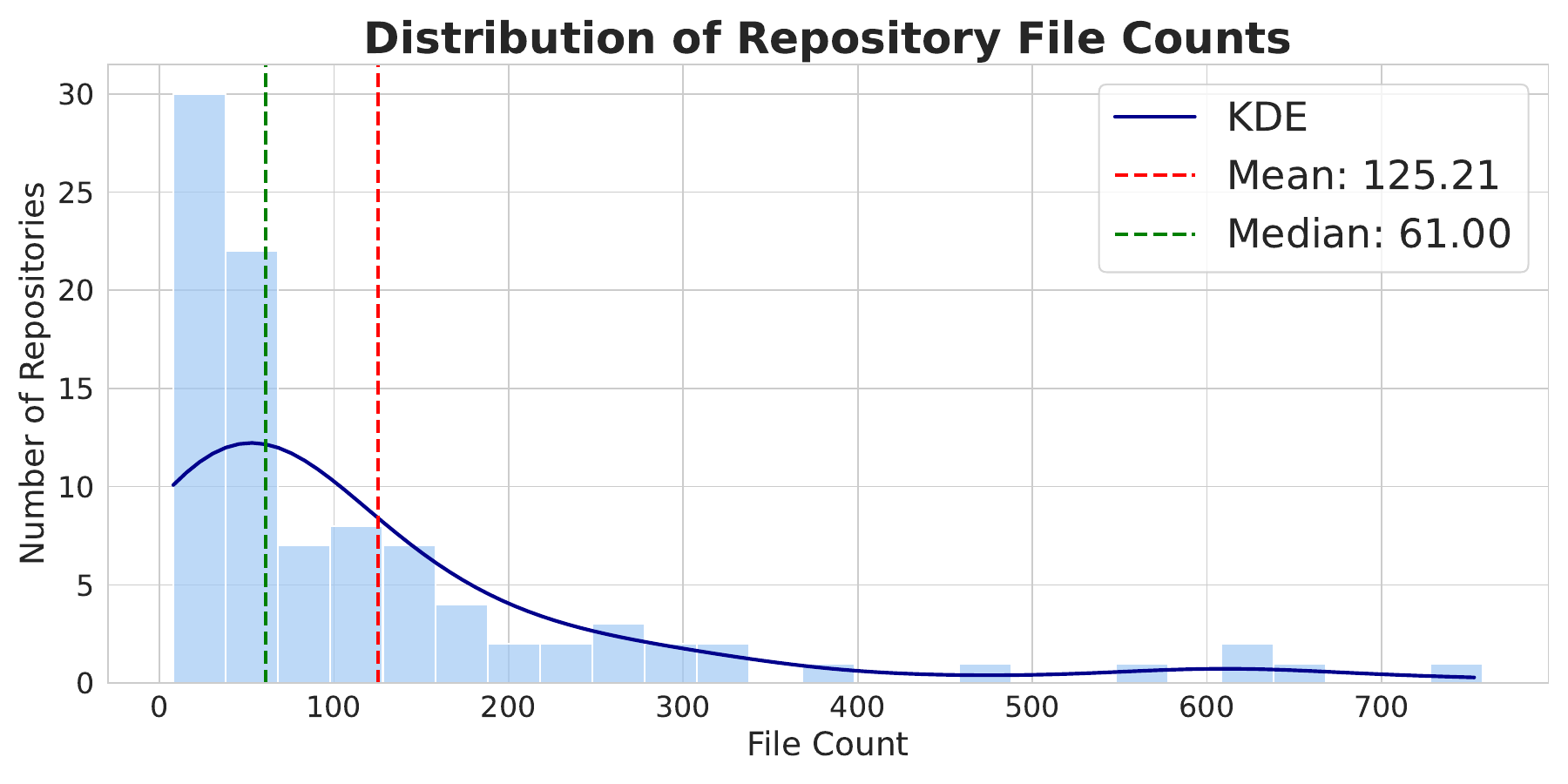}
    \vspace{-1.5em}
    \caption{Number of Files per Repository}
    \label{fig:file_count_distribution}
\end{figure}

Figure \ref{fig:file_count_distribution} depicts the distribution of file counts in repositories, reflecting their complexity and scale based on the number of files. Because the number of files in most repositories are in the lower hundreds, it is feasible to leverage directory structure for code deployment with LLMs.

\begin{figure}[ht]
    \centering
    \includegraphics[width=\linewidth]{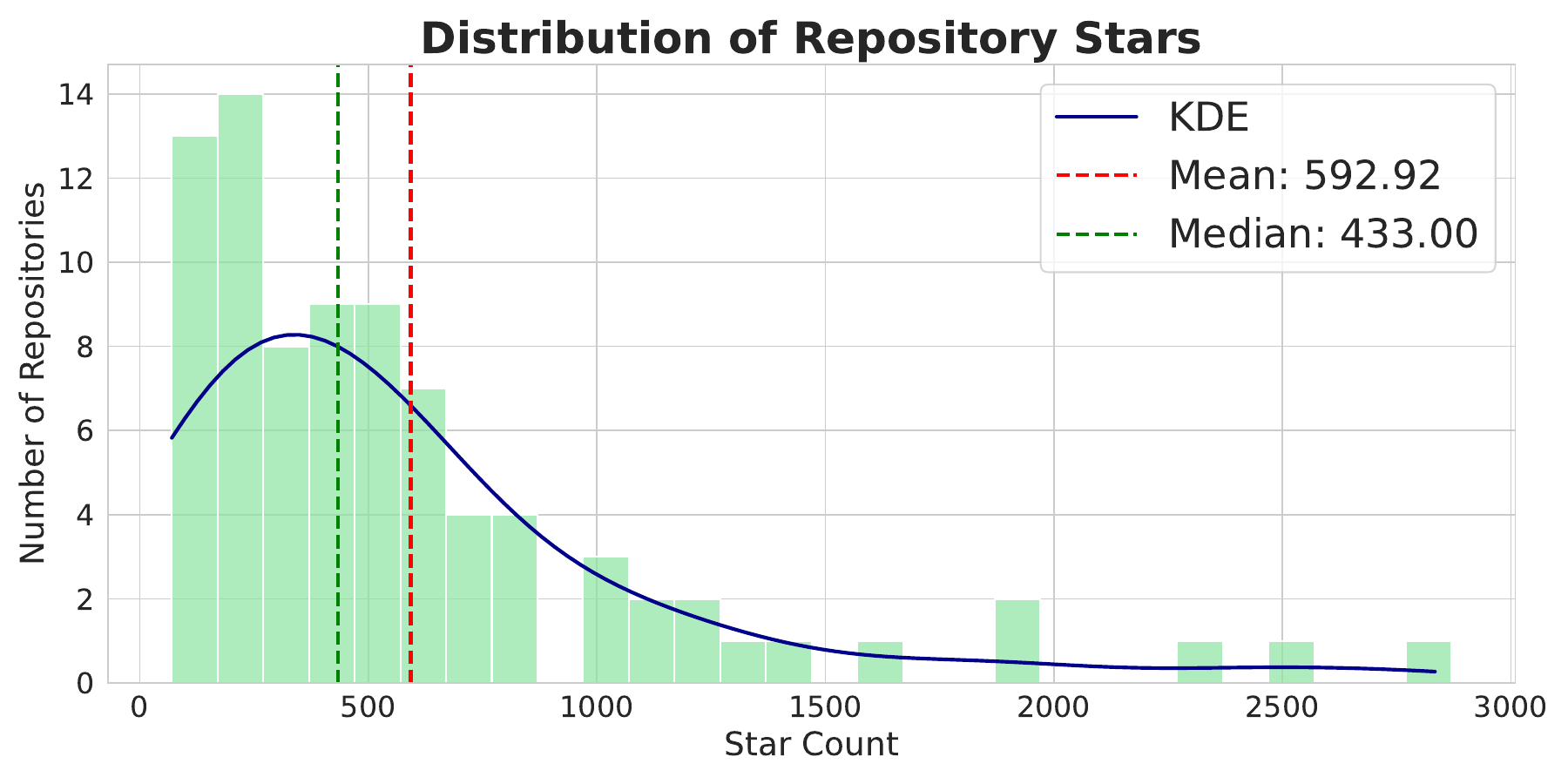}
    \vspace{-1.5em}
    \caption{Stargazer Distribution}
    \label{fig:stars_distribution}
\end{figure}

Figure \ref{fig:stars_distribution} shows the distribution of star counts in selected repositories. The average count is over 590, which means that these repositories receives significant attention and indicates they are generally well maintained.

\begin{figure}[ht]
    \centering
    \includegraphics[width=\linewidth]{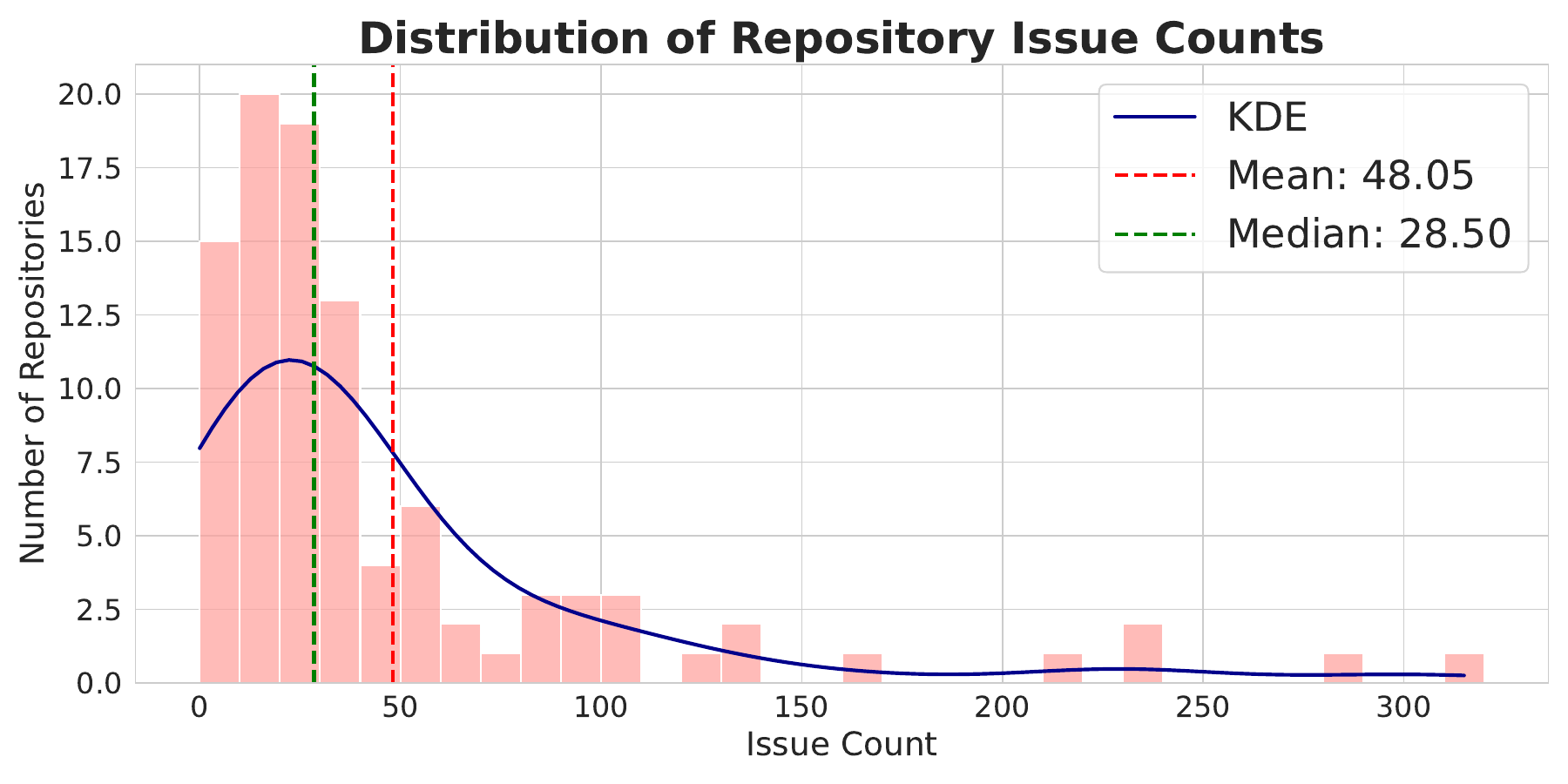}
    \vspace{-1.5em}
    \caption{Size of Issue Database}
    \label{fig:issue_count_distribution}
\end{figure}

Figure \ref{fig:issue_count_distribution} shows the distribution of issues counts in selected repositories. On average, each repository contains over 48 issues, indicating these repositories have a good amount of engagements with the open-source community and sufficient support from the authors. Therefore, the information in the issues are valuable for the deployment tasks.


\section{CSR-Agents}

In this section, we propose \agent, a multi-agent framework that leverages LLMs for different tasks and achieves effective cooperation among agents for code deployment tasks. We introduce our standardized environment for code deployment, functionalities of different agents, and their cooperation workflow.

\subsection{Code Deployment with LLM}
For each repository, we use the README file under the root folder as the major source of information and prompt LLMs to generate executable bash commands for different steps of deployment including environment setup, data preparation, training, inference and evaluation. 


To achieve reproducbility and safety, we use the Docker container to isolate the code deployment environment. We build a standard Docker image that is equipped with essential tools like bash, Conda, GCC, Make, Python, etc. for evaluations across all repositories and various LLMs. In the container, we use a counter to count the total number of scripts that needed to be executed and the number of scripts that were executed successfully. In this way, we safeguard the entire computing system, especially from bash command executions that involve system-level permissions and could potentially break the whole system.

We note that the use of Docker is not part of the original repositories. The introduction of Docker into our evaluation system can ensure that each time we conduct the evaluation, we start from the same environment for fair comparisons across different LLMs, and the evaluation of one repository does not affect any other repositories. Another benefit of using Docker is that we can speed up the evaluation by running different evaluations in separate docker instances at the same time.

The LLM generated bash commands may not work well for successful deployment due to various reasons. First, in some README files, some basic directives (such as \verb|conda| and \verb|pip|) were missing, so the LLMs could not generate these commands at the first attempt. Secondly, the installation of packages may not be completed in a single attempt and usually needs several iterations of trial-and-error. Thirdly, some steps of the deplopyment require checking additional information like GitHub issues of the corresponding repositories and from the internet. To handle these issues, we design a group of LLM agents that cooperate effectively.

\subsection{\raisebox{-12pt}{\includegraphics[height=30pt]{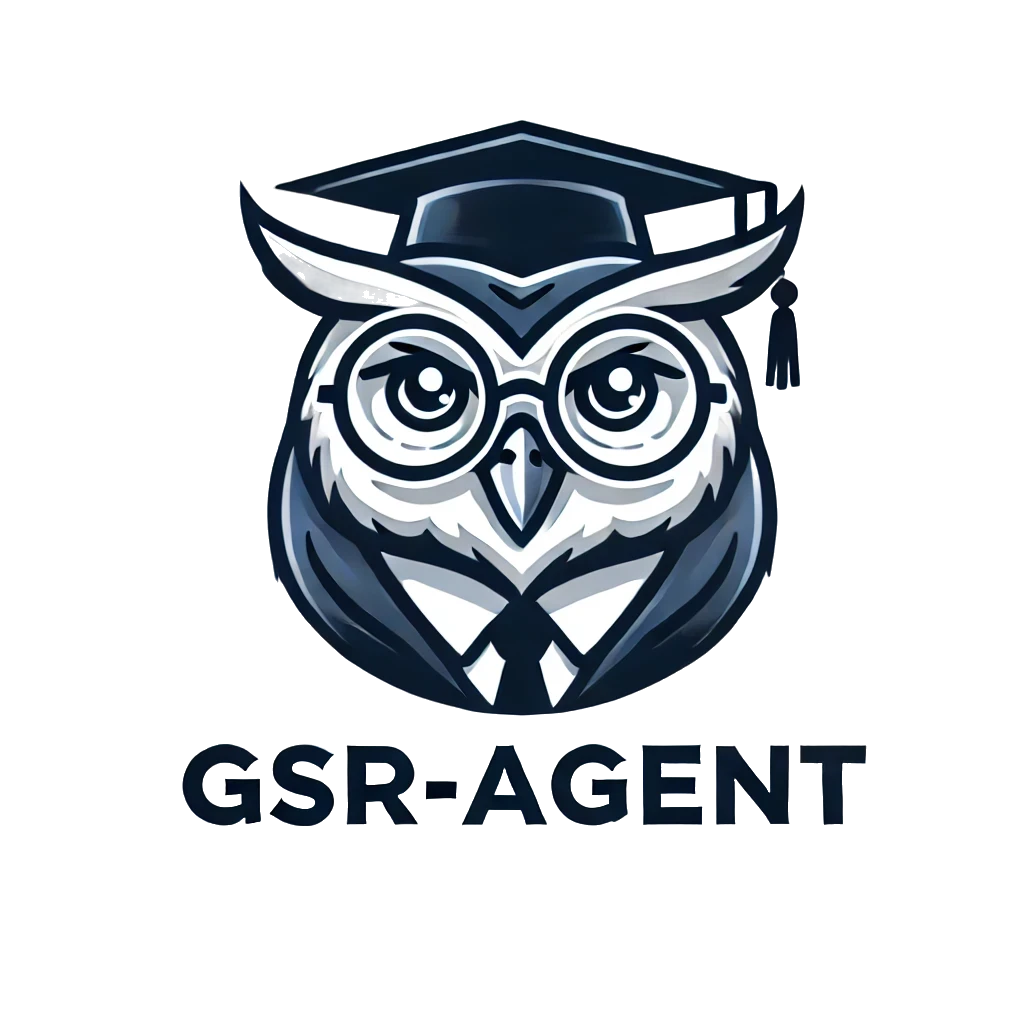}} \hspace{-0.5em} CSR-Agent: LLM Agent Design}

\begin{figure*}[h]
    \centering
    \includegraphics[width=\linewidth]{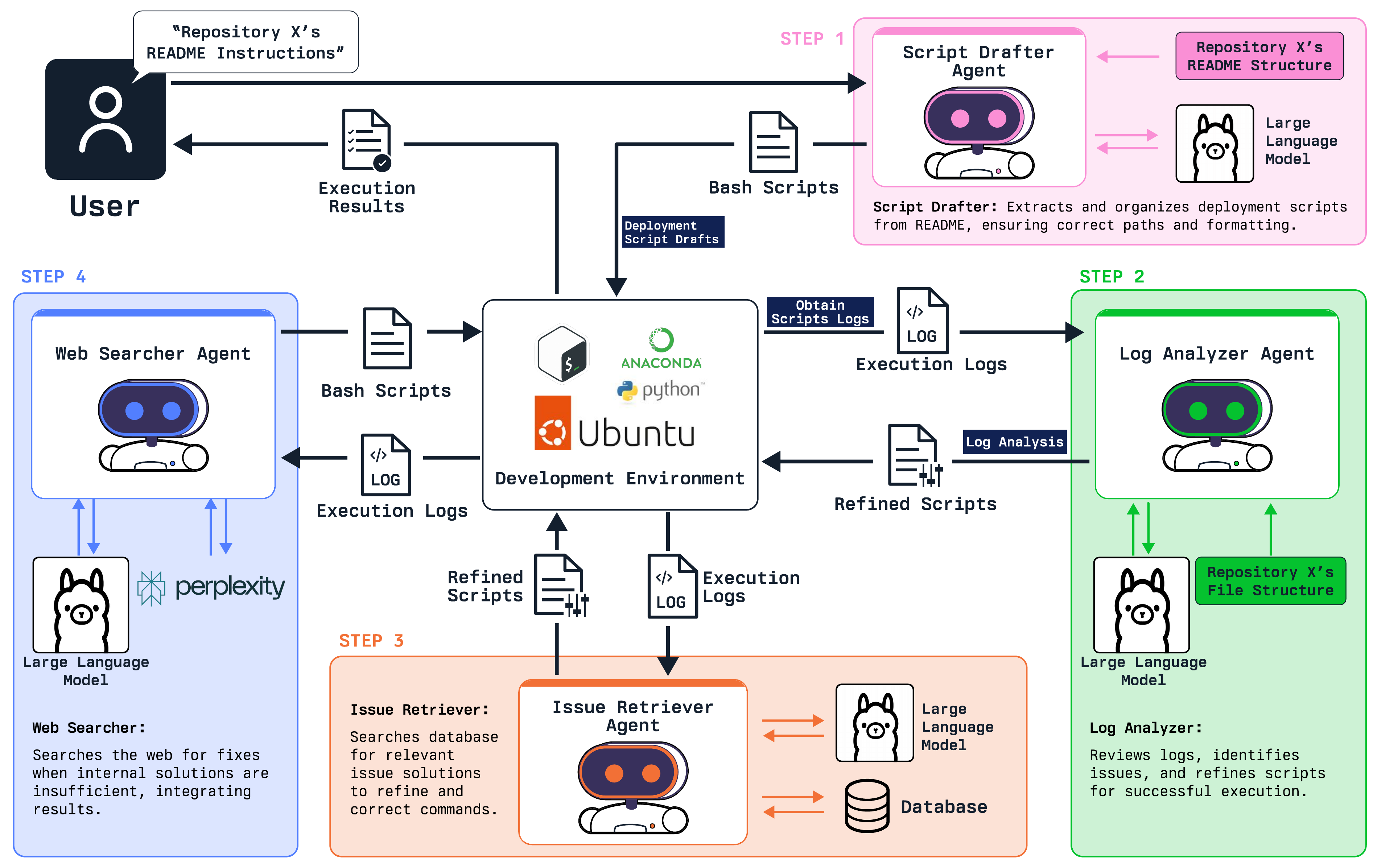}
    \caption{Workflow of \model}
    \label{fig:workflow}
\end{figure*}

In \agent, we adopt an iterative trial-and-error process for successful code deployment. Specifically, LLM takes README, directory structure, and error logs from failed command execution as input and generates bash commands to complete the deployment tasks.
The system comprises five agents: Command Drafter, Script Executor, Log Analyzer, Issue Retriver, and Web Searcher. These agents collectively facilitate the deployment of a code repository. The complete workflow is shown in Figure \ref{fig:workflow}.

\textbf{Command Drafter}: This agent reads the README files and directory structure and generates a draft script containing bash commands for deployment. Then, it divides the entire script into five sections, each corresponding to a step in the code repository deployment. This sectional division also serves as an evaluation standard later on.

\textbf{Script Executor}: This agent receives the draft commands from the Command Drafter and execute the commands in our standardized Docker environment. After execution, it collects logs from \verb|bash|, including standard output and errors. Note that during bash script execution, no explicit return code is provided by \verb|bash|. We experimented with setting predefined special prompt to \verb|bash| and parse the return code from the returned output and error message. However, we find that a good quantity of commands do not have return code, making the feedback from \verb|bash| not informative to other agents.  To address this challenge, we leverage the LLM in the executor, instructing it to provide feedback based on the standard output and error messages. We then parse this feedback to generate a return code: if the return code is zero, then the command is executed successfully; otherwise, the command, output, and error messages are logged and sent to the Log Analyzer.


\textbf{Log Analyzer}: This agent reads the logs and the associated bash command, and checks for updates, missing prerequisites, or script paths that need updating. It also identifies any other missing components and returns a curated command for a successful execution. We note that this agent only reply on the internal knowledge of the LLM and may not be able to correctly solve the errors.

\textbf{Issue Retriever}: This agent takes in the command, standard output and error message and search them against the issue database we collected from the repository. It leverages RAG and its pipeline requires a search algorithm to query the input against the database. In our design, the queries are the combination of the commands executed, standard output and error messages. The database contains GitHub issues and their communication records for the repository. We experimented with BM25 and Contriever as the retrieval algorithm and decided to use BM25 for (1) BM25's higher search speed and (2) the fact that error logs and issues generally share keywords, so sentimental search do not possess much advantage over lexical search.

\textbf{Web Searcher}: This agent utilizes the Perplexity API to obtain solutions for failed execution. If the pipeline reaches this stage, it indicates that the Log Analyzer and Issue Retriever failed to solve a failed command. The standard output, standard error, and the failed command are fed to Perplexity to search the web for solutions. The Web Searcher then analyzes the solutions from Perplexity and generates new bash commands to resolve the issue.


\subsection{LLM Coorporation Framwfork}
The workflow operates as follows: Deployment commands are drafted from repository documentation, executed in a bash environment, and adjusted based on log analysis if errors arise. Additional information is retrieved from an issue database and web search if needed. The process concludes with a summary that outlines successes and identifies steps needing further attention.

\parabf{Drafting the Initial Commands} The instructions from each repository's README are fed into the Command Drafter LLM Agent to draft the necessary commands for deploying the repository. These commands are organized into five stages: prerequisite installation, data and model checkpoints downloading, training, inference, and evaluation. It is important to note that not every repository contains all these sections, and some sections may be empty.

\parabf{Execution of Commands} After the draft stage, the script drafted will be sent to the Script Executor. To provide LLM Agents with the Python interface, we implement a BashExecutor that encapsulates the bash binary executable file in it. If the commands execute successfully, the process is deemed successful and the deployment pipeline will return \verb|True|. Though looks simple, the bash simulation was not straightforward during our experiments, since we need to consider the environment variables, stdout, stderr, etc. Initially, we used a subprocess to handle it, but the success rate was extremely low. Upon analyzing the logs, we discovered that no commands related to environment variable changes were reflected in subsequent instructions. For instance, if the Command changes the environment to a Python virtual environment or another \verb|conda| environment, this change only applies to that specific command. Afterward, the Python interpreter and package manager revert to their default settings. Besides, not all command executions have a valid return code, therefore, we utilize an LLM to parse the standard output and standard error (if any) to obtain the return state of the execution.

\parabf{Analyze the Execution Log} If the execution is not successful and error occurs, the standard output, standard error, command, and return code are logged and analyzed. The Log Analyzer examines the error messages and attempts to refine the execution commands or adjust prerequisites to ensure the environment is prepared. We have a \verb|max_attempt| argument that limits the number of retries for the log analyzer. If the issue persists after certain attempts, the workflow utilizes the web search tool to request external information.

\parabf{Retrieve Augmented Generation from Issue Database}
After analyzing the execution log and making initial adjustments, the next step is to retrieve more insights using a Retrieve Augmented Generation (RAG) approach. The Issue Retriever agent uses the logged command, output, and error messages to query the issue database for similar past problems or discussions. Leveraging the BM25 retrieval algorithm, it matches keywords from the logs to relevant entries. If a match is found, the agent extracts solutions or troubleshooting steps, feeding them back into the workflow to refine commands. If no relevant match is found, the process escalates to the Web Searcher for external information.


\parabf{Search the Internet for External Information} Using tools such as Perplexity, the WebSearcher agent integrates the command error logs and standard output with external information. This process refines the command and retries execution in the bash with a limited number of attempts. The agent records the information and transfers it to the deployment summarizer if the issue remains unresolved.

\section{Evaluations}

In this section, we evaluate \agent in \model with a wide range of popular foundation LLM families, including Claude~\footnote{https://www.anthropic.com/news/claude-3-family}, GPT-4~\cite{achiam2023gpt}, Llama-3~\cite{dubey2024llama}, and Mistral~\footnote{https://mistral.ai/technology/\#models}. For each LLM family, we experimented with different model sizes for thorough comparisons.

We show the completion rate in Table~\ref{tab:initial_success}, Table~\ref{tab:analyzer_success}, Table~\ref{tab:ragger_success}, and Table~\ref{tab:searcher_success}. In these tables, we use \textbf{S} to stand for the Setup stage, \textbf{D} to stand for the ownload stage, \textbf{T} to stand for the Training stage, \textbf{E} to stand for the Evaluation stage, and \textbf{I} for the inference stage. 



\subsection{Initial Drafter} 
As shown in Table~\ref{tab:initial_success}, all models perform well on \textit{Setup} and \textit{Download} tasks (success rates around 0.23 to 0.28) but struggle with \textit{Training}, \textit{Evaluation}, and \textit{Inference}, where success rates are close to zero. This indicates that the drafter agent handles basic installation effectively but has difficulty with complex tasks requiring updates to file paths and environment variables.

We note that this is also similar to the deployment process of a real researcher, where their first execution is more likely to fail and they need to analyze the errors and leverage tools like GitHub issues or search engines to solve the problem.

\begin{table}[htbp]
  \centering
  \caption{Drafter Success Metrics of Different Models.}
  \resizebox{\linewidth}{!}{%
  \begin{tabular}{ll ccccc}
  \toprule
  \textbf{Model Type} & \textbf{Name} & \textbf{S} & \textbf{D} & \textbf{T} & \textbf{E} & \textbf{I} \\
  \midrule
  \multirow[=]{3}{*}{\textbf{Claude}} 
  & Instant & 0.232 & 0.189 & 0.007 & 0.000 & 0.000 \\
  & 3-Haiku & 0.253 & 0.239 & 0.046 & 0.005 & 0.052 \\
  & 3-Sonnet & 0.284 & 0.283 & 0.045 & 0.024 & 0.031 \\
  \midrule
  \multirow[=]{3}{*}{\textbf{GPT}} 
  & 4o-Mini & 0.242 & 0.229 & 0.008 & 0.016 & 0.029 \\
  & 4o & 0.261 & 0.238 & 0.039 & 0.022 & 0.031 \\
  & 4-Turbo & 0.271 & 0.252 & 0.028 & 0.039 & 0.032 \\
  \midrule
  \multirow[=]{4}{*}{\textbf{Llama}} 
  & 3-70B & 0.239 & 0.306 & 0.019 & 0.040 & 0.032 \\
  & 3.1-8B & 0.243 & 0.200 & 0.051 & 0.037 & 0.007 \\
  & 3.1-70B & 0.260 & 0.280 & 0.032 & 0.022 & 0.019 \\
  \midrule
  \multirow[=]{2}{*}{\textbf{Mistral}} 
  & Large & 0.243 & 0.266 & 0.047 & 0.031 & 0.026 \\
  & Large-2 & 0.251 & 0.279 & 0.039 & 0.025 & 0.024 \\
  \bottomrule
  \end{tabular}%
  }
  \label{tab:initial_success}
\end{table}

\subsection{Log Analyzer} 
Table~\ref{tab:analyzer_success} shows noticeable improvements in all tasks compared to the drafter stage. Success rates for \textit{Setup} and \textit{Download} increase to around 0.34 to 0.40, while complex tasks see gains up to 0.18.

\begin{table}[htbp]
  \centering
  \caption{Analyzer Success Metrics of Different Models.}

  \resizebox{\linewidth}{!}{%
  \begin{tabular}{ll ccccc}
  \toprule
  \textbf{Model Type} & \textbf{Name} & \textbf{S} & \textbf{D} & \textbf{T} & \textbf{E} & \textbf{I} \\
  \midrule
  \multirow{3}{*}{\textbf{Claude}} 
  & Instant & 0.342 & 0.353 & 0.104 & 0.109 & 0.151 \\
  & 3-Haiku & 0.350 & 0.301 & 0.132 & 0.037 & 0.130 \\
  & 3-Sonnet & 0.388 & 0.400 & 0.168 & 0.116 & 0.129 \\
  \midrule
  \multirow{3}{*}{\textbf{GPT}} 
  & 4o-Mini & 0.347 & 0.317 & 0.118 & 0.078 & 0.131 \\
  & 4o & 0.362 & 0.353 & 0.148 & 0.115 & 0.145 \\
  & 4-Turbo & 0.353 & 0.322 & 0.161 & 0.094 & 0.148 \\
  \midrule
  \multirow{4}{*}{\textbf{Llama}} 
  & 3-70B & 0.361 & 0.382 & 0.111 & 0.185 & 0.176 \\
  & 3.1-8B & 0.304 & 0.386 & 0.183 & 0.114 & 0.123 \\
  & 3.1-70B & 0.313 & 0.335 & 0.141 & 0.182 & 0.151 \\
  \midrule
  \multirow{2}{*}{\textbf{Mistral}} 
  & Large & 0.324 & 0.349 & 0.121 & 0.143 & 0.144 \\
  & Large-2 & 0.340 & 0.357 & 0.152 & 0.199 & 0.163 \\
  \bottomrule
  \end{tabular}%
  }

  \label{tab:analyzer_success}
\end{table}

Analyzers leverage dynamic feedback from the execution of commands to refine scripts and try to correct errors from executed commands.

\subsection{Issue Retriever} 
In the Issue Retriever stage (Table~\ref{tab:ragger_success}), success rates continue to improve, especially for complex tasks like \textbf{Training}, \textbf{Evaluation}, and \textbf{Inference}, reaching up to 0.25.

\begin{table}[tbp]
  \centering
  \caption{Issue Retriever Success Metrics of Different Models.}
  \resizebox{\linewidth}{!}{%
  \begin{tabular}{ll ccccc}
  \toprule
  \textbf{Model Type} & \textbf{Name} & \textbf{S} & \textbf{D} & \textbf{T} & \textbf{E} & \textbf{I} \\
  \midrule
  \multirow{3}{*}{\textbf{Claude}} 
  & Instant & 0.365 & 0.369 & 0.129 & 0.130 & 0.169 \\
  & 3-Haiku & 0.374 & 0.329 & 0.139 & 0.061 & 0.143 \\
  & 3-Sonnet & 0.442 & 0.436 & 0.254 & 0.183 & 0.163 \\
  \midrule
  \multirow{3}{*}{\textbf{GPT}} 
  & 4o-Mini & 0.375 & 0.367 & 0.171 & 0.122 & 0.159 \\
  & 4o & 0.379 & 0.375 & 0.169 & 0.128 & 0.160 \\
  & 4-Turbo & 0.377 & 0.381 & 0.178 & 0.126 & 0.164 \\
  \midrule
  \multirow{4}{*}{\textbf{Llama}} 
  & 3-70B & 0.364 & 0.399 & 0.113 & 0.159 & 0.154 \\
  & 3.1-8B & 0.305 & 0.389 & 0.182 & 0.115 & 0.122 \\
  & 3.1-70B & 0.312 & 0.334 & 0.143 & 0.179 & 0.153 \\
  \midrule
  \multirow{2}{*}{\textbf{Mistral}} 
  & Large & 0.357 & 0.358 & 0.174 & 0.155 & 0.153 \\
  & Large-2 & 0.359 & 0.380 & 0.181 & 0.155 & 0.152 \\
  \bottomrule
  \end{tabular}%
  }
  \label{tab:ragger_success}
\end{table}

The results show that access to a knowledge base with informative discussions on issues of the code repository allows LLMs to retrieve solutions to execution errors from earlier stages, enhancing performance in complex operations.

\subsection{Web Searcher}
The Searcher Success Metrics in Table~\ref{tab:searcher_success} exhibits the highest performance. Success rates for \textit{Setup} and \textit{Download} reach up to 0.46, and complex tasks improve to between 0.15 and 0.29.

\begin{table}[htbp]
  \centering
  \caption{Searcher Success Metrics of Different Models.}
  \resizebox{\linewidth}{!}{%
  \begin{tabular}{ll ccccc}
  \toprule
  \textbf{Model Type} & \textbf{Name} & \textbf{S} & \textbf{D} & \textbf{T} & \textbf{E} & \textbf{I} \\
  \midrule
  \multirow{3}{*}{\textbf{Claude}} 
  & Instant & 0.388 & 0.406 & 0.151 & 0.131 & 0.190 \\
  & 3-Haiku & 0.385 & 0.338 & 0.155 & 0.070 & 0.163 \\
  & 3-Sonnet & 0.467 & 0.467 & 0.291 & 0.194 & 0.189 \\
  \midrule
  \multirow{3}{*}{\textbf{GPT}} 
  & 4o-Mini & 0.412 & 0.405 & 0.201 & 0.131 & 0.179 \\
  & 4o & 0.415 & 0.407 & 0.198 & 0.130 & 0.183 \\
  & 4-Turbo & 0.416 & 0.406 & 0.200 & 0.133 & 0.182 \\
  \midrule
  \multirow{4}{*}{\textbf{Llama}} 
  & 3-70B & 0.380 & 0.442 & 0.200 & 0.173 & 0.170 \\
  & 3.1-8B & 0.318 & 0.447 & 0.201 & 0.157 & 0.165 \\
  & 3.1-70B & 0.344 & 0.463 & 0.183 & 0.196 & 0.184 \\
  \midrule
  \multirow{2}{*}{\textbf{Mistral}} 
  & Mistral-Large & 0.375 & 0.450 & 0.199 & 0.174 & 0.168 \\
  & Mistral-Large-2 & 0.373 & 0.452 & 0.201 & 0.172 & 0.169 \\
  \bottomrule
  \end{tabular}%
  }
  \label{tab:searcher_success}
  \vspace{-0.5em}
\end{table}

Web search enables models to find up-to-date solutions, resolving issues that previous agents could not, leading to substantial improvements.



\subsection{Aggregated Results}

\begin{figure*}[htbp]
    \centering
    \begin{minipage}[b]{0.48\linewidth}
        \centering
        \includegraphics[width=\linewidth]{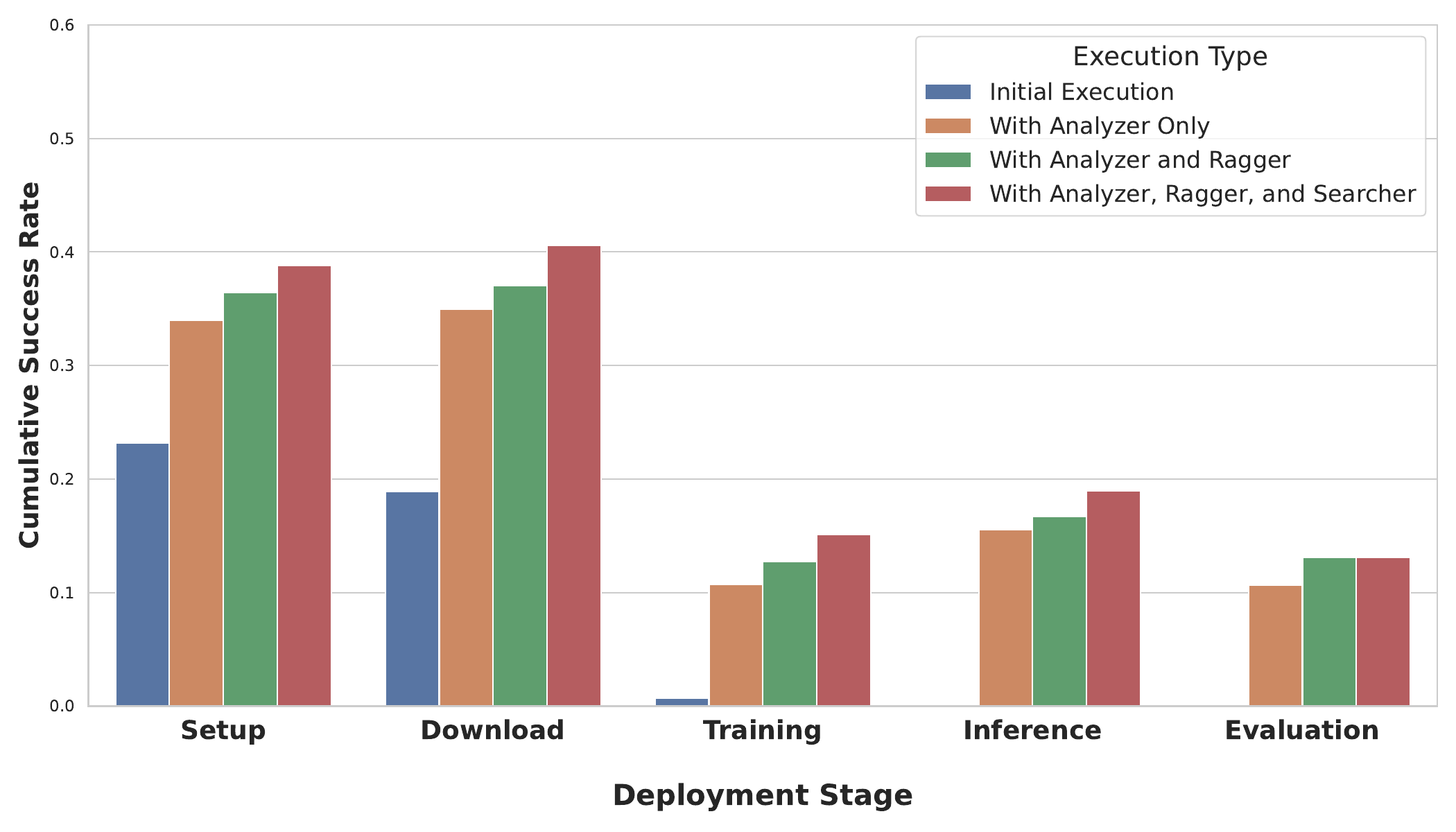}
        \caption{Performance of Claude 3 Sonnet}
        \label{fig:claude}
    \end{minipage}
    \hfill
    \begin{minipage}[b]{0.48\linewidth}
        \centering
        \includegraphics[width=\linewidth]{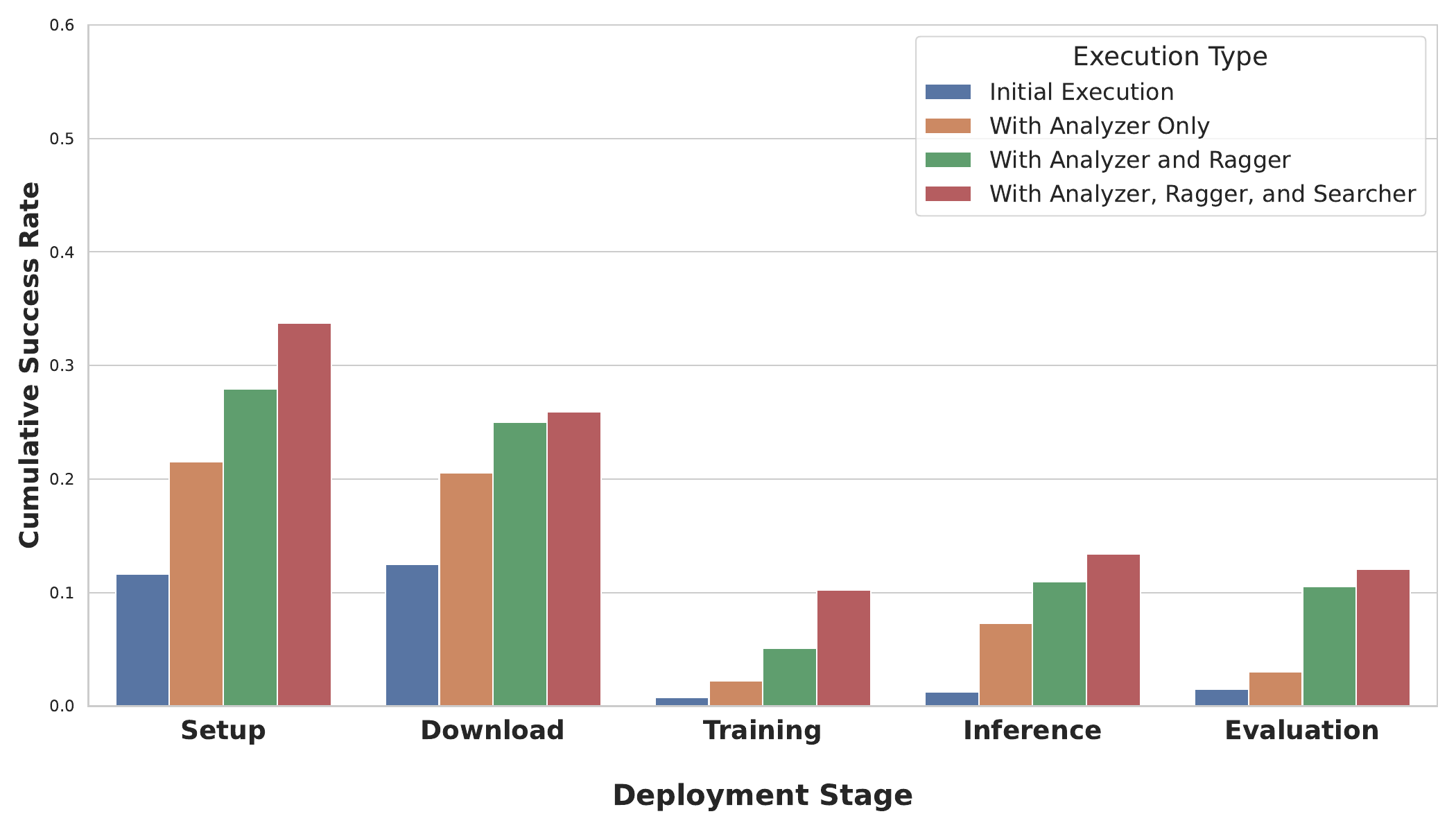}
        \caption{Performance of GPT 4o}
        \label{fig:gpt}
    \end{minipage}

\end{figure*}

\begin{figure*}[htbp]
    \centering
    \begin{minipage}[b]{0.48\linewidth}
        \centering
        \includegraphics[width=\linewidth]{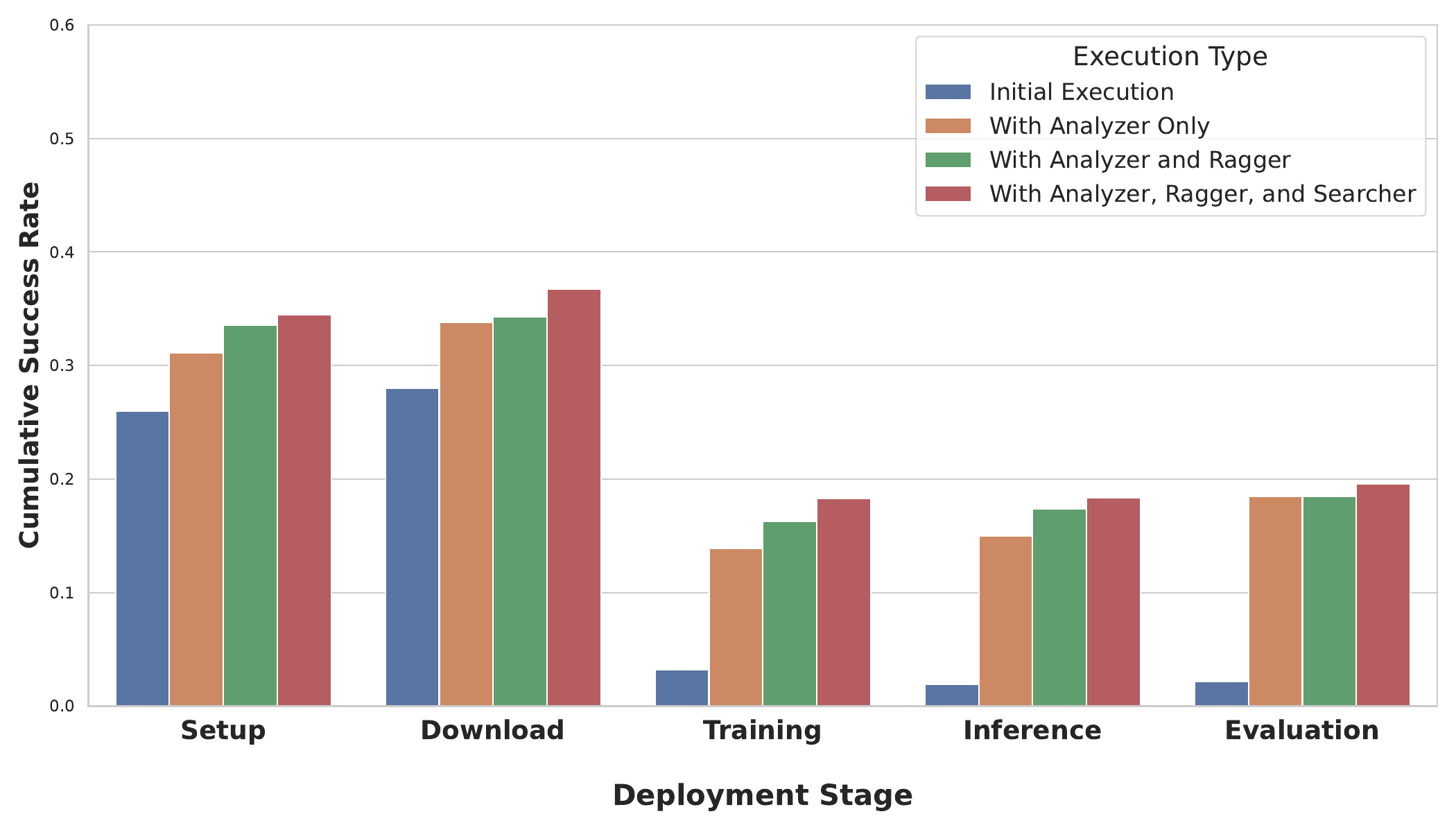}
        \caption{Performance of Llama 3.1 70B Instruct}
        \label{fig:llama}
    \end{minipage}
    \hfill
    \begin{minipage}[b]{0.48\linewidth}
        \centering
        \includegraphics[width=\linewidth]{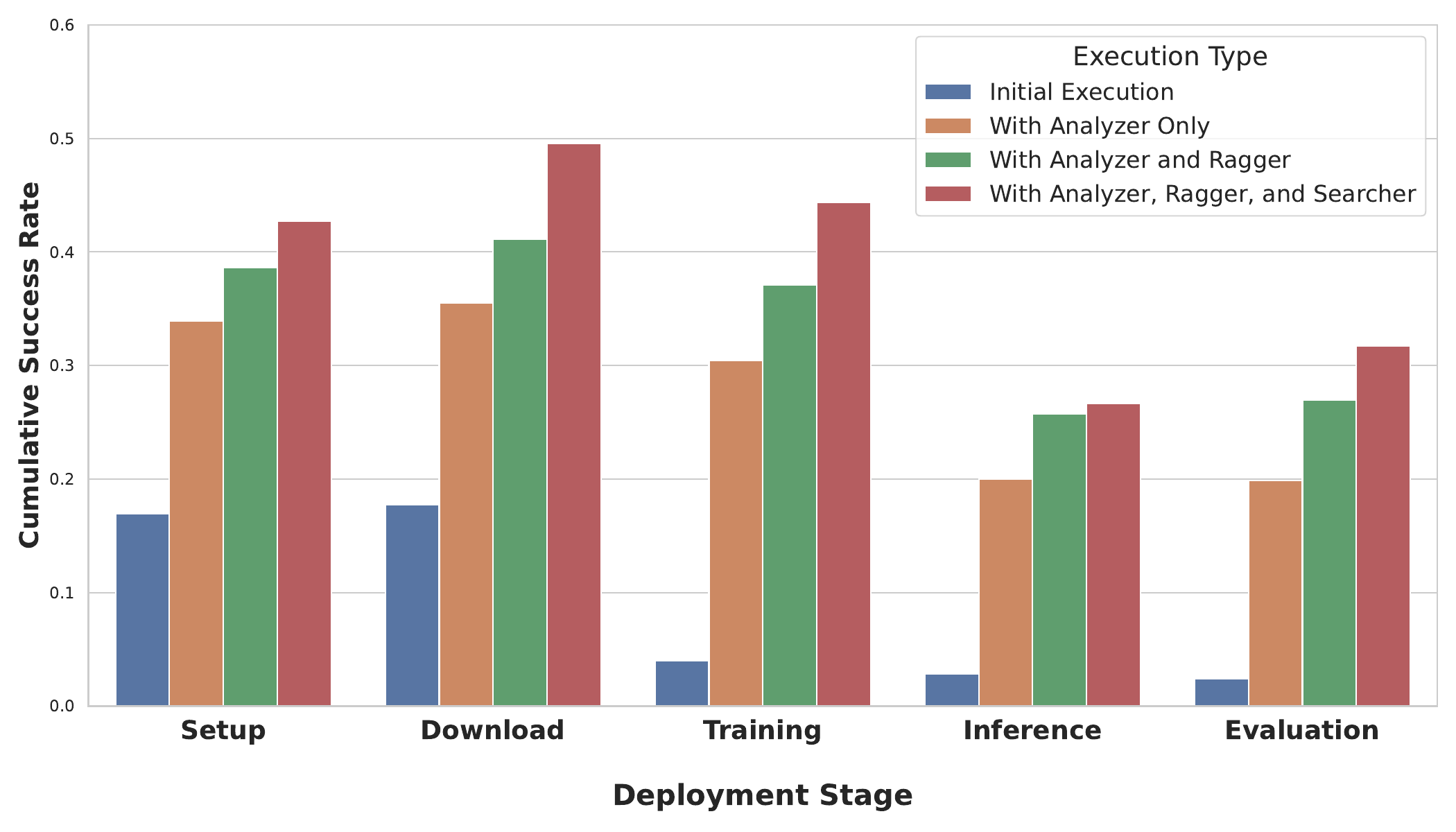}
        \caption{Performance of Mistral Large 2}
        \label{fig:mistral}
    \end{minipage}

\end{figure*}

We show the aggregated results of a single LLM on different tasks with different level of engagements of multi-agents in Figure~\ref{fig:claude}, Figure~\ref{fig:gpt}, Figure~\ref{fig:llama}, and Figure~\ref{fig:mistral}. In short, with more agents contributing to solving the tasks, the success rate increases across all tasks for all LLMs, which demonstrates the effectiveness of our proposed \agent.
\subsection{Results Interpretation} \label{results}
The evaluation of the \model involved assessing the performance of various large language models (LLMs) across key tasks: Setup, Download, Training, Inference, and Evaluation. These tasks are essential for deploying repositories within the \model.

\subsubsection{Task-Specific Outcomes}
\parabf{Setup and Download:} Most models consistently performed well, reflecting their capability to initiate and manage basic deployment processes.

\parabf{Inference and Evaluation:} Performance was less consistent, with some models demonstrating moderate success, but generally struggling with the complexity of these tasks.

\parabf{Training:} Training tasks are particularly challenging, with lower success rates across the board, indicating that current LLMs require further refinement to handle training processes effectively.

\subsubsection{Overall Performance}

The success metrics across different tasks and models indicate a wide variability in performance. Generally, models showed higher success rates in Setup and Download tasks, with performance tapering off in more complex tasks such as Inference, Evaluation, and Training. This pattern highlights the challenges LLMs face in handling the full deployment process autonomously.

The results demonstrate that while LLMs have made significant strides in automating repository deployment, their ability to manage complex tasks remains limited. Improvements are needed, particularly in the areas of Inference and Training, to achieve fully autonomous and reliable deployment of science repositories.

However, there is still a large gap between LLMs and real scientists even if the advnanced tools are provided to the LLMs. To explain, it is not trivial to handle the nuances in the experiement environment setup for the science repositories. For example, the hardware and software compatibility issues are very common in code deployment and often causes confusions even for domain experts.


\section{Conclusion} \label{conclusion}
\vspace{-0.5em}

The work introduces \model, a benchmark designed to evaluate the capabilities of LLM agents in automating the deployment of GitHub repositories for scientific research. Our study highlights that while LLMs show potential in handling tasks like environment setup and data preparation, they face challenges in complex tasks such as training and inference, where success rates are notably lower.

Our multi-agent framework, \agent, exemplifies how LLMs can collaborate to tackle deployment challenges, offering a promising approach to improving automation in software engineering. However, the results indicate that further advancements are needed to fully realize autonomous and reliable deployment processes.

Overall, \model serves as a crucial tool for assessing and improving LLM-driven deployment workflows in scientific research, paving the way for more efficient and automated computer science projects exploration.

\section*{Limitations}
Although our benchmark framework supports several tools to facilitate large language model agents in the code deployment task, it does not actually improve the original reasoning capabilities of the large language models that are used in the agents. To improve LLMs' reasoning capabilities for this specific task, the community may resort to techniques like RLHF, which is orthogonal to this work. Our benchmark only focuses on code repositories that related to computer science research topics, and does not involve other types of repositories. Although this framework can be reused for other types of the repositories, we do not explore that direction in this work, and leave it to future works.
\newpage
\bibliography{custom}
\clearpage
\newpage
\appendix

\section{Appendix}

\subsection{Categories and Subcategories}

\subsection*{Computer Vision and Graphics}
\begin{itemize}
    \item \textbf{3D Vision and Reconstruction}
    \begin{itemize}
        \item Structured3D: Structured 3D modeling dataset for computer vision and graphics (ECCV 2020)
        \item RayDF: Light ray-based 3D rendering and learning (NeurIPS 2023)
        \item SSDNeRF: Diffusion-based NeRF for 3D reconstruction (ICCV 2023)
        \item NeuralPull: Learning distance functions for point clouds (ICML 2021)
        \item DAD-3DHeads: Dataset for 3D head alignment and reconstruction (CVPR 2022)
        \item LeGO-LOAM: Lidar-based odometry and mapping for robotics
    \end{itemize}
    \item \textbf{Image Segmentation, Processing, and Enhancement}
    \begin{itemize}
        \item Unsupervised-Semantic-Segmentation: Unsupervised methods for semantic segmentation (ICCV 2021)
        \item Seg-Uncertainty: Scene adaptation for self-driving cars (IJCAI 2020, IJCV 2021)
        \item RGBD\_Semantic\_Segmentation\_PyTorch: RGBD scene recognition and segmentation (ECCV 2020)
        \item BCNet: Occlusion-aware segmentation (CVPR 2021)
        \item Deep\_Metric: Metric learning for image retrieval
        \item Multiview2Novelview: Novel view synthesis from multiple views (ECCV 2018)
        \item NeRCo: Low-light image enhancement using neural representations (ICCV 2023)
        \item Transfiner: High-quality instance segmentation with mask transformers (CVPR 2022)
    \end{itemize}
    \item \textbf{Pose Estimation and Object Detection}
    \begin{itemize}
        \item HybrIK: Hybrid kinematics for 3D human pose (CVPR 2021)
        \item SmoothNet: Human pose refinement in videos (ECCV 2022)
        \item FSA-Net: Head pose estimation from images (CVPR 2019)
        \item VisualDet3D: Monocular 3D detection for autonomous driving
        \item DEVIANT: Monocular 3D object detection for autonomous driving (ECCV 2022)
        \item OWOD: Open World Object Detection with incremental learning (CVPR 2021)
    \end{itemize}
\end{itemize}

\subsection*{Machine Learning Models and Techniques}
\begin{itemize}
    \item \textbf{Generative and Diffusion Models}
    \begin{itemize}
        \item DDM2: Diffusion models for MRI denoising (ICLR 2023)
        \item DDNM: Diffusion models for zero-shot image restoration (ICLR 2023)
        \item diffusion-point-cloud: Generating 3D point clouds with diffusion models (CVPR 2021)
        \item GeoDiff: Diffusion models for molecular structures (ICLR 2022)
        \item probabilistic\_unet: Conditional segmentation with probabilistic UNet (NeurIPS 2018)
        \item LTSF-Linear: Linear models for time series forecasting (AAAI 2023)
        \item unconditional-time-series-diffusion: Diffusion models for time series prediction (NeurIPS 2023)
        \item dyffusion: Spatiotemporal forecasting using diffusion models (NeurIPS 2023)
    \end{itemize}
    \item \textbf{Graph Neural Networks and Transformers}
    \begin{itemize}
        \item GraphSAINT: Scalable graph neural networks with sampling (ICLR 2020, IPDPS 2019)
        \item Graph-Transformer: Applying transformers to graph data (WWW 2022)
        \item graphtransformer: Generalized transformers for graphs (DLG-AAAI 2021)
        \item GPT-GNN: Pre-training techniques for graph neural networks (KDD 2020)
        \item GraphMAE: Masked autoencoders for graph learning (KDD 2022)
    \end{itemize}
    \item \textbf{Self-Supervised and Contrastive Learning}
    \begin{itemize}
        \item skip-connections-matter: Impact of skip connections on adversarial example transferability (ICLR 2020)
        \item SparK: BERT-style pretraining on convolutional networks (ICLR 2023)
        \item imbalanced-semi-self: Semi-supervised learning for class-imbalanced datasets (NeurIPS 2020)
        \item CURL: Image enhancement via neural curve layers (ICPR 2020)
    \end{itemize}
    \item \textbf{Model Interpretability and Optimization}
    \begin{itemize}
        \item rrl: Rule-based learning for interpretable classification (NeurIPS 2021)
        \item gradient-descent-the-ultimate-optimizer: Advanced optimization techniques (NeurIPS 2022)
    \end{itemize}
\end{itemize}

\subsection*{Large Language Models (LLMs)}
\begin{itemize}
    \item \textbf{LLM Development and Optimization}
    \begin{itemize}
        \item llama-recipes: Fine-tuning scripts for LLaMA models
        \item llama3: Official GitHub repository for LLaMA 3
        \item alpaca-lora: Instruct-tuning LLaMA on consumer hardware
        \item MetaGPT: Multi-agent AI development framework using LLMs
        \item Stanford\_Alpaca: Training and dataset generation for Alpaca models
    \end{itemize}
    \item \textbf{Inference and Efficiency in LLMs}
    \begin{itemize}
        \item mistral-inference: Efficient inference for Mistral models
        \item PowerInfer: Fast LLM serving on consumer-grade GPUs
        \item direct-preference-optimization: Implementation of direct preference optimization (DPO)
        \item GPTCache: Efficient caching for LLMs integrated with LangChain and llama\_index
    \end{itemize}
    \item \textbf{LLM Applications and Evaluation}
    \begin{itemize}
        \item alpaca\_eval: Automatic evaluation for instruction-following LLMs
        \item DART: Enhancing few-shot learning with differentiable prompts (ICLR 2022)
        \item LRV-Instruction: Reducing hallucination in multi-modal models (ICLR 2024)
        \item alpaca\_farm: Simulated framework for reinforcement learning from human feedback (RLHF)
        \item FastChat: Open-source platform for training and evaluating large language models
    \end{itemize}
\end{itemize}

\begin{figure*}[h]
    \centering
    \includegraphics[width=0.9\linewidth]{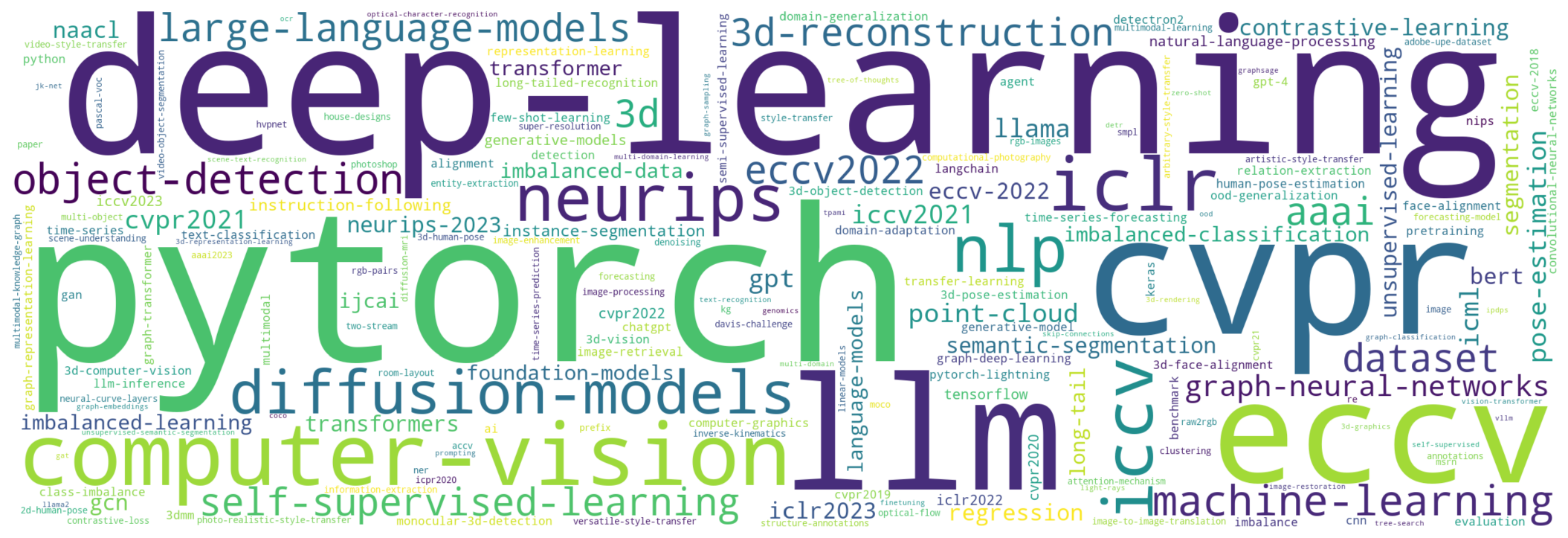}
    \caption{Distribution of Topics in GSRBench100}
    \label{fig:topics_distribution}
\end{figure*}

\begin{table*}[ht]
\small
\centering
\caption{Categories, Subcategories, and Abbreviations}
\label{tab:categories_subcategories_abbreviations}
\begin{tabular}{cll}
\toprule
\textbf{Category} & \textbf{Subcategory} & \textbf{Subcategory Abbreviation} \\
\midrule
\multirow{3}{*}{\textbf{CV}} 
    & 3D Vision \& Reconstruction & 3D Vision \& Rec. \\
    & Image Segmentation \& Enhancement & Img Segm. \& Enh. \\
    & Pose Estimation \& Object Detection & Pose Estim. \& Obj Det. \\
\midrule
\multirow{4}{*}{\textbf{ML}} 
    & Generative \& Diffusion Models & Gen. \& Diff. Models \\
    & Graph Neural Networks \& Transformers & GNNs \& Transf. \\
    & Self-Supervised \& Contrastive Learning & Self-Sup. \& Contr. Learn. \\
    & Model Interpretability \& Optimization & Model Interp. \& Opt. \\
\midrule
\multirow{3}{*}{\textbf{NLP}} 
    & Text Classification \& Extraction & Txt Classif. \& Extr. \\
    & Multimodal \& Knowledge Systems & Multi. \& Knowl. Sys. \\
    & Transfer Learning \& Domain Adaptation & Transf. Learn. \& Dom. Adap. \\
\midrule
\multirow{3}{*}{\textbf{LLM}} 
    & LLM Development \& Optimization & LLM Dev. \& Opt. \\
    & Inference \& Efficiency in LLMs & Infer. \& Eff. in LLMs \\
    & LLM Applications \& Evaluation & LLM Appl. \& Eval. \\
\midrule
\multirow{3}{*}{\textbf{Interdisciplinary}} 
    & Healthcare \& Bioinformatics & Health. \& Bioinf. \\
    & Robotics \& Autonomous Systems & Robotics \& Auto. Sys. \\
    & Computational Creativity \& Arts & Comp. Creativ. \& Arts \\
\bottomrule
\end{tabular}
\end{table*}

\vspace{-1em}
\subsection*{Natural Language Processing (NLP)}
\begin{itemize}
    \item \textbf{Text Classification and Information Extraction}
    \begin{itemize}
        \item HVPNeT: Multimodal entity and relation extraction (NAACL 2022)
        \item PURE: Simple approach to relation extraction (NAACL 2021)
        \item SimCSE: Contrastive sentence embeddings (EMNLP 2021)
        \item Text-GCN: Text classification using graph convolutional networks (AAAI 2019)
    \end{itemize}
    \item \textbf{Multimodal and Knowledge-Based Systems}
    \begin{itemize}
        \item OntoProtein: Protein function prediction using knowledge graphs (ICLR 2022)
        \item tree-of-thought-llm: Structured problem-solving with LLMs (NeurIPS 2023)
        \item storm: Automated report generation using large language models
        \item SWE-agent: AI-based code bug fixing using LLMs
    \end{itemize}
    \item \textbf{Transfer Learning and Domain Adaptation}
    \begin{itemize}
        \item SubpopBench: Benchmark for subpopulation shifts in domain generalization (ICML 2023)
        \item naacl\_transfer\_learning\_tutorial: Transfer learning techniques in NLP (NAACL 2019)
        \item private-transformers: Training transformers with differential privacy
        \item imbalanced-regression: Handling data imbalance in regression tasks (ICML 2021)
    \end{itemize}
\end{itemize}

\subsection*{Domain-Specific Applications}
\begin{itemize}
    \item \textbf{Healthcare and Bioinformatics}
    \begin{itemize}
        \item imbalanced-regression: Addressing imbalances in healthcare data (ICML 2021)
        \item hyena-dna: Long-range genomic modeling with Hyena
        \item OntoProtein: Protein function prediction with pretraining (ICLR 2022)
        \item GeoDiff: Molecular structure generation using diffusion models (ICLR 2022)
        \item tape: Benchmark tasks for protein sequence modeling
    \end{itemize}
    \item \textbf{Robotics and Autonomous Systems}
    \begin{itemize}
        \item LeGO-LOAM: Lidar-based odometry and mapping for robotics
        \item SuperGlobal: Image retrieval using global features (ICCV 2023)
        \item Deep\_Metric: Embedding learning for metric-based image retrieval
        \item FEARTracker: Robust visual tracking (ECCV 2022)
    \end{itemize}
    \item \textbf{Computational Creativity and Visual Arts}
    \begin{itemize}
        \item CCPL: Artistic style transfer for images and videos (ECCV 2022)
        \item GANgealing: GAN-based visual alignment (CVPR 2022)
        \item NeRCo: Low-light image enhancement using neural representations (ICCV 2023)
    \end{itemize}
\end{itemize}

\subsection{Word Cloud of Topics}
Figure \ref{fig:topics_distribution} presents a word cloud that visualizes the frequency of topics across the repositories, highlighting the primary focus areas within GSRBench100. It provides a quick overview of the thematic concentration of the dataset.

\vspace{-0.5em}
\section{Repository list and Commit ID}

We specify the commit ID we used in our benchmark dataset for reproducibility. Please see table~\ref{tab:repo_Part 1} and table~\ref{tab:repo_Part 2} for details.

\vspace{-0.5em}
\section{Categories, Subcategories, and Abbreviations}

\begin{table*}[htbp]
  \caption{GitHub Repository Details - Part 1}
  \label{tab:repo_Part 1}
  \resizebox{\linewidth}{!}{%
  \begin{tabular}{|l|l|l|}
    \hline
    Repository URL & Commit ID & Branch \\
    \hline
    \url{https://github.com/cleardusk/3DDFA\_V2} & 1b6c67601abffc1e9f248b291708aef0e43b55ae & master \\
\hline
\url{https://github.com/lkeab/BCNet} & d6580e8a2a0b5e71c0ae6913ed0340c101d35723 & main \\
\hline
\url{https://github.com/JarrentWu1031/CCPL} & d0b6b326d7d624b5e8d1543a3a84a745a08fd883 & main \\
\hline
\url{https://github.com/sjmoran/CURL} & 4be9753a8063f9833423e4aa5947ae7a64b114f8 & master \\
\hline
\url{https://github.com/tiangexiang/CurveNet} & c2e7cf642b7e08d9aec5b70263f6989a85c9e191 & main \\
\hline
\url{https://github.com/PinataFarms/DAD-3DHeads} & 3acc5c2a1177d354a1247c49e44a83ad682ea6a1 & main \\
\hline
\url{https://github.com/zjunlp/DART} & d418ded8bd548ef25f2d030990e707a497e93483 & main \\
\hline
\url{https://github.com/StanfordMIMI/DDM2} & d07be20ad36446f8e35621d4b0d92e7cf54c169e & main \\
\hline
\url{https://github.com/wyhuai/DDNM} & 00b58eac7843a4c99114fd8fa42da7aa2b6808af & main \\
\hline
\url{https://github.com/abhi1kumar/DEVIANT} & 009955f3bbb21c38a687eaae59bdfcb82eca93e7 & main \\
\hline
\url{https://github.com/shenweichen/DSIN} & e8ba406eeda0916214897d44866bffc419c3edb0 & master \\
\hline
\url{https://github.com/bnu-wangxun/Deep\_Metric} & 04ca51093db13135a04e3c94401bc898c6af0c40 & master \\
\hline
\url{https://github.com/tjiiv-cprg/EPro-PnP} & 42412220b641aef9e8943ceba516b3175631d370 & main \\
\hline
\url{https://github.com/PinataFarms/FEARTracker} & 0a3bd039918909c79c1b7e55a4bfb7807520abde & main \\
\hline
\url{https://github.com/shamangary/FSA-Net} & 4361d0e48103bb215d15734220c9d17e6812bb48 & master \\
\hline
\url{https://github.com/lm-sys/FastChat} & 92a6d1fcd69a88ea169c0b01065ce44f1e690a2c & main \\
\hline
\url{https://github.com/acbull/GPT-GNN} & f26e13c69ddc8a3f2580cb16d0b9a1c73d89f4bc & master \\
\hline
\url{https://github.com/zilliztech/GPTCache} & 75ab7ec7b871c8399a95d5bf528441f2856250dd & main \\
\hline
\url{https://github.com/MinkaiXu/GeoDiff} & ea0ca48045a2f7abfccd7f0df449e45eb6eae638 & main \\
\hline
\url{https://github.com/daiquocnguyen/Graph-Transformer} & 99c88a116148fdaa8d3071fcc548e5c471ae607f & master \\
\hline
\url{https://github.com/THUDM/GraphMAE} & b14f080c919257b495e3cb6474286d5252d6a635 & main \\
\hline
\url{https://github.com/GraphSAINT/GraphSAINT} & c9b1e340d7b951465ac4a9251eef93832e68b003 & master \\
\hline
\url{https://github.com/zjunlp/HVPNeT} & 52c77f7835a295d9c8534997b1316c42b2662972 & main \\
\hline
\url{https://github.com/IDEA-Research/HumanSD} & c5db29dd66a3e40afa8b4bed630f0aa7ea001880 & main \\
\hline
\url{https://github.com/Jeff-sjtu/HybrIK} & 9b8681dcf3c902dd5dacc01520ba04982990e1e2 & main \\
\hline
\url{https://github.com/princeton-nlp/LM-BFF} & c282f521001f9c299d29eec7b459266f2b14fbaf & main \\
\hline
\url{https://github.com/FuxiaoLiu/LRV-Instruction} & 0a5ab538ed96c3e0c9835b5fe02cc8f7fa0bf8fa & main \\
\hline
\url{https://github.com/cure-lab/LTSF-Linear} & 0c113668a3b88c4c4ee586b8c5ec3e539c4de5a6 & main \\
\hline
\url{https://github.com/RobustFieldAutonomyLab/LeGO-LOAM} & 896a7a95a8bc510b76819d4cc48707e344bad621 & master \\
\hline
\url{https://github.com/tfzhou/MATNet} & c8b95e527c486c304f711cc7dffb060f31abe19f & master \\
\hline
\url{https://github.com/MIVRC/MSRN-PyTorch} & a0e038de7eb42e21d2e88c38e6490b61a02c566e & master \\
\hline
\url{https://github.com/princeton-nlp/MeZO} & 552cb1b710767f9a6e1dc8f9645d7640376f9941 & main \\
\hline
\url{https://github.com/geekan/MetaGPT} & 5446c7e490e7203c61b2ff31181551b2c0f4a86b & main \\
\hline
\url{https://github.com/shaohua0116/Multiview2Novelview} & a5e236f3c564bf287c8a09d855fd2134ba86b299 & master \\
\hline
\url{https://github.com/Ysz2022/NeRCo} & 6b0e1112231d0902976ad76357044de582a307f3 & main \\
\hline
\url{https://github.com/mabaorui/NeuralPull} & c093a52308a9b74446d24cc6c1b0fee5ee5bb7bb & master \\
\hline
\url{https://github.com/JosephKJ/OWOD} & 23890f188cd1a6801c6ac0e3dacd78b8572b8c29 & master \\
\hline
\url{https://github.com/zjunlp/OntoProtein} & 6360f458e11670ecfaf853ee68f2087b31439dc0 & main \\
\hline
\url{https://github.com/princeton-nlp/PURE} & b1e9cad39bec10eb3c355dc5a8e4e75dd0afebf5 & main \\
\hline
\url{https://github.com/SJTU-IPADS/PowerInfer} & 61cac9bf25e60336bbad27ada9dbb809204473ac & main \\
\hline
\url{https://github.com/charlesCXK/RGBD\_Semantic\_Segmentation\_PyTorch} & 32b3f86822d278103a13ea6f93f9668d3b631398 & master \\
\hline
\url{https://github.com/vLAR-group/RayDF} & ca6c663523b777732788a5d8100d36251a482b31 & master \\
\hline
\url{https://github.com/Paranioar/SGRAF} & 50d0c6f9caf759099b28371046f780342357c405 & main \\
\hline
\url{https://github.com/Lakonik/SSDNeRF} & b9d195db76bb715c475b24287362d9627d77d3bb & main \\
\hline
\url{https://github.com/shamangary/SSR-Net} & f98b6cbe1c9c8c78649e5a331f94113564521525 & master \\
\hline
\url{https://github.com/princeton-nlp/SWE-agent} & 36e430d27ffd11269738df92d6c521cab2207dcb & main \\
\hline
\url{https://github.com/daveredrum/ScanRefer} & 9d7483053e8d29acfd4db4eb1bc28f1564f5dddb & master \\
\hline
\url{https://github.com/layumi/Seg-Uncertainty} & 6fce9eae141c2c0592b3e7c1b3e5f8ee7b1ce9a6 & master \\
\hline
\url{https://github.com/princeton-nlp/SimCSE} & 7edb07e05cec0d5293fc1696b578d8056dba76ef & main \\
\hline
\url{https://github.com/cure-lab/SmoothNet} & c03e93e8a14f55b9aa087dced2751a7a5e2d50b0 & main \\
\hline

  \end{tabular}
  }
\end{table*}

\begin{table*}[htbp]
  \caption{GitHub Repository Details - Part 2}
  \label{tab:repo_Part 2}
  \resizebox{\linewidth}{!}{%
  \begin{tabular}{|l|l|l|}
    \hline
    Repository URL & Commit ID & Branch \\
    \hline
    \url{https://github.com/keyu-tian/SparK} & a63e386f8e5186bc07ad7fce86e06b08f48a61ea & main \\
\hline
\url{https://github.com/bertjiazheng/Structured3D} & d06d1b241b290ae7ed6b311d710ffafffcb567de & master \\
\hline
\url{https://github.com/YyzHarry/SubpopBench} & 4d3dbbe21029666ef19d040e110ec22908640c5b & main \\
\hline
\url{https://github.com/ShihaoShao-GH/SuperGlobal} & 86946964b907e6f28a7264add2c15640fae30009 & main \\
\hline
\url{https://github.com/codeKgu/Text-GCN} & 70b970a52efcb80235cf0ae3e578eaf80278d5f3 & master \\
\hline
\url{https://github.com/wvangansbeke/Unsupervised-Semantic-Segmentation} & dfd5fa0a1542f2b26824b4059bd2bb1240c5c94b & main \\
\hline
\url{https://github.com/tloen/alpaca-lora} & 8bb8579e403dc78e37fe81ffbb253c413007323f & main \\
\hline
\url{https://github.com/tatsu-lab/alpaca\_eval} & 32c8c0d068205c38b02003b67e0beec82a8f1ac2 & main \\
\hline
\url{https://github.com/tatsu-lab/alpaca\_farm} & 30717ddae735365de756ee2085191b491a71788d & main \\
\hline
\url{https://github.com/adobe/antialiased-cnns} & b27a34a26f3ab039113d44d83c54d0428598ac9c & master \\
\hline
\url{https://github.com/richardaecn/class-balanced-loss} & 1d7857208a2abc03d84e35a9d5383af8225d4b4d & master \\
\hline
\url{https://github.com/luost26/diffusion-point-cloud} & 1e30d48d018820fbc7c67c8b3190215bd41878e4 & main \\
\hline
\url{https://github.com/eric-mitchell/direct-preference-optimization} & f8b8c0f49dc92a430bae41585f9d467d3618fe2f & main \\
\hline
\url{https://github.com/Rose-STL-Lab/dyffusion} & 832574f6f788a0cd4a4d75e8f59b3c07c7e8446b & main \\
\hline
\url{https://github.com/wpeebles/gangealing} & ffa6387c7ffd3f7de76bdc693dc2272e274e9bfd & main \\
\hline
\url{https://github.com/openai/gpt-3} & d7a9bb505df6f630f9bab3b30c889e52f22eb9ea & master \\
\hline
\url{https://github.com/kach/gradient-descent-the-ultimate-optimizer} & b3b047e02ca6d32e0e61e34a0ca6e0bc57e55bdf & main \\
\hline
\url{https://github.com/microsoft/graphrag} & 61b5eea34783c58074b3c53f1689ad8a5ba6b6ee & main \\
\hline
\url{https://github.com/graphdeeplearning/graphtransformer} & c9cd49368eed4507f9ae92a137d90a7a9d7efc3a & main \\
\hline
\url{https://github.com/HazyResearch/hyena-dna} & d553021b483b82980aa4b868b37ec2d4332e198a & main \\
\hline
\url{https://github.com/YyzHarry/imbalanced-regression} & a6fdc45d45c04e6f5c40f43925bc66e580911084 & main \\
\hline
\url{https://github.com/YyzHarry/imbalanced-semi-self} & b91ad29fd8805ddf0a146f735905b0c869e68ae4 & master \\
\hline
\url{https://github.com/meta-llama/llama} & 8fac8befd776bc03242fe7bc2236cdb41b6c609c & main \\
\hline
\url{https://github.com/meta-llama/llama-recipes} & 8c1418e93b817cb6734a9cfe095b270f5a0f48f5 & main \\
\hline
\url{https://github.com/meta-llama/llama3} & 18f515a3c3c5f02cf45c6ac56cc5d039488e867a & main \\
\hline
\url{https://github.com/state-spaces/mamba} & a71bb5a83bfa289b5807aefc1767232dee77b35e & main \\
\hline
\url{https://github.com/kwotsin/mimicry} & a7fda06c4aff1e6af8dc4c4a35ed6636e434c766 & master \\
\hline
\url{https://github.com/mistralai/mistral-finetune} & 5b8adb54a1263664d52dab6f94581bf24d7b59e3 & main \\
\hline
\url{https://github.com/mistralai/mistral-inference} & 1f583071dc7aad2ca35cb9896140316ffece5b65 & main \\
\hline
\url{https://github.com/YyzHarry/multi-domain-imbalance} & 2efbfefd34542e365293f798d79f70cee5e54303 & main \\
\hline
\url{https://github.com/JiangWenPL/multiperson} & e8ae029cc691f3f9c3958a23f762f3d72cf65c54 & master \\
\hline
\url{https://github.com/zju3dv/mvpose} & 38b958f423f2de2bf7562f5a386c27440eab8c53 & master \\
\hline
\url{https://github.com/huggingface/naacl\_transfer\_learning\_tutorial} & dc976775bb11edee24a77e2ce161450089c5e169 & master \\
\hline
\url{https://github.com/baudm/parseq} & 1902db043c029a7e03a3818c616c06600af574be & main \\
\hline
\url{https://github.com/lxuechen/private-transformers} & 18ccc4eab7355e4ac96051a82434796f6aa4624b & main \\
\hline
\url{https://github.com/SimonKohl/probabilistic\_unet} & 7a2e79d549184d0f3a47d0deaa054a70b0f54a3f & master \\
\hline
\url{https://github.com/Jeff-sjtu/res-loglikelihood-regression} & 203dc3195ee5a11ed6f47c066ffdb83247511359 & master \\
\hline
\url{https://github.com/12wang3/rrl} & f8d0886b23c4e15f63c62c248b97d4eb73386ad1 & main \\
\hline
\url{https://github.com/csdongxian/skip-connections-matter} & 9b2e5cca9b673efcac253e16b2f55f6cda1a8692 & master \\
\hline
\url{https://github.com/xuchen-ethz/snarf} & ae0c893cc049f0f8270eaa401e138dff5d4637b9 & main \\
\hline
\url{https://github.com/tatsu-lab/stanford\_alpaca} & 761dc5bfbdeeffa89b8bff5d038781a4055f796a & main \\
\hline
\url{https://github.com/stanford-oval/storm} & 263d894e34e2a399a18f80fbf3c9028dc06e1a1d & main \\
\hline
\url{https://github.com/songlab-cal/tape} & 6d345c2b2bbf52cd32cf179325c222afd92aec7e & master \\
\hline
\url{https://github.com/SysCV/transfiner} & 5b61fb53d8df5484f44c8b7d8415f398fd283ddc & main \\
\hline
\url{https://github.com/yzhq97/transmomo.pytorch} & 0e4d2f0e7e8af159e018c8a85f9d789de963a83a & master \\
\hline
\url{https://github.com/princeton-nlp/tree-of-thought-llm} & ab400345c5ea39d28ea6d7d3be0e417b11113c87 & master \\
\hline
\url{https://github.com/amazon-science/unconditional-time-series-diffusion} & 3eafeffdffefd4d2f5344ac63e5627a746c331a3 & main \\
\hline
\url{https://github.com/elliottwu/unsup3d} & dc961410d61684561f19525c2f7e9ee6f4dacb91 & master \\
\hline
\url{https://github.com/dddzg/up-detr} & 44c0c6eb4bddf409a41cbaae31b7360062495199 & master \\
\hline
\url{https://github.com/Owen-Liuyuxuan/visualDet3D} & ad229f2e491cba2c48f9bb2c211fe672294e2398 & master \\
\hline
  \end{tabular}
  }
\end{table*}

\subsubsection{Example Text from README}

Below are case studies of how models (GPT-4o, Claude 3, LLaMA 3.1, and Mistral) extract the commands from a Repository.


\begin{table*}[th!]
\fontsize{9.0pt}{\baselineskip}\selectfont
\linespread{0.9}\selectfont
\begin{mybody}
\textbf{REPOSITORY ID:} 3DDFA\_V2 \\\\
\textbf{README} \begin{lstlisting}[language=bash]
Requirements
See requirements.txt, tested on macOS and Linux platforms. The Windows users 
may refer to FQA for building issues. Note that this repo uses Python3. The 
major dependencies are PyTorch, numpy, opencv-python, and onnxruntime, etc. 
If you run the demos with --onnx flag to do acceleration, you may need to 
install libomp first, i.e., brew install libomp on macOS.

Usage

# 1. Clone this repo
git clone https://github.com/cleardusk/3DDFA_V2.git
cd 3DDFA_V2

# 2. Build the cython version
sh ./build.sh

# 3. Run demos
# Running on a still image
python3 demo.py -f examples/inputs/emma.jpg --onnx

# Running on videos
python3 demo_video.py -f examples/inputs/videos/214.avi --onnx

# Running on webcam
python3 demo_webcam_smooth.py --onnx

The implementation of tracking is simply by alignment. If the head pose > 90 
degrees or the motion is too fast, the alignment may fail. A threshold is 
used to trickly check the tracking state, but it is unstable.
\end{lstlisting}
\end{mybody}
\begin{mybody}
\textbf{GPT-4o:}\\
Extracted commands for Environment Setup / Requirement / Installation:
\begin{verbatim}
git clone https://github.com/cleardusk/3DDFA_V2.git; cd 3DDFA_V2; sh ./build.sh
\end{verbatim}

Error message from bash shell:
\begin{verbatim}
"stderr": "Traceback (most recent call last):\n  File \"/workspace/3DDFA_V2/FaceBoxes/
utils/build.py\", line 15, in <module>\n    from Cython.Distutils import build_ext\n
ModuleNotFoundError: No module named 'Cython'\nTraceback (most recent call last):\n  
File \"/workspace/3DDFA_V2/Sim3DR/setup.py\", line 7, in <module>\n    from Cython.
Build import cythonize\nModuleNotFoundError: No module named 'Cython'",

Analyzer suggested solution:
"command": "git clone https://github.com/cleardusk/3DDFA_V2.git; cd 3DDFA_V2; 
pip install cython; sh ./build.sh"
\end{verbatim}
\end{mybody}
\captionof{table}{Commands and errors for setting up 3DDFA\_V2 repository using GPT-4o.}
\label{tab:commands_1}
\end{table*}

\clearpage 
\newpage 

\begin{table*}[th!]
\fontsize{9.0pt}{\baselineskip}\selectfont
\linespread{0.9}\selectfont

\begin{mybody}
\textbf{LLaMA 3.1:}\\
Extracted commands for Environment Setup / Requirement / Installation:
\begin{verbatim}
pip install -r requirements.txt
brew install libomp
\end{verbatim}
\end{mybody}

\begin{mybody}
\textbf{Claude 3:}\\
Extracted commands for Environment Setup / Requirement / Installation:
\begin{verbatim}
git clone https://github.com/cleardusk/3DDFA_V2.git
cd 3DDFA_V2
sh ./build.sh
\end{verbatim}

Error message from bash shell:
\begin{verbatim}
"stderr": "Traceback (most recent call last):\n  File \"/workspace/3DDFA_V2/FaceBoxes/
utils/build.py\", line 15, in <module>\n    from Cython.Distutils import build_ext\n
ModuleNotFoundError: No module named 'Cython'\nTraceback (most recent call last):\n  
File \"/workspace/3DDFA_V2/Sim3DR/setup.py\", line 7, in <module>\n    from Cython.
Build import cythonize\nModuleNotFoundError: No module named 'Cython'",

Analyzer suggested solution:
"command": "pip install Cython && sh ./build.sh",
\end{verbatim}
\end{mybody}

\begin{mybody}
\textbf{Mistral:}\\
Extracted commands for Environment Setup / Requirement / Installation:
\begin{verbatim}
git clone https://github.com/cleardusk/3DDFA_V2.git && cd 3DDFA_V2 && sh ./build.sh &&
brew install libomp
\end{verbatim}

Error message from bash shell:
\begin{verbatim}
"stderr": "bash: brew: command not found"

Analyzer suggested solution:
"command": "sudo apt-get install homebrew && git clone https://github.com/cleardusk/
3DDFA_V2.git && cd 3DDFA_V2 && sh ./build.sh && brew install libomp",
\end{verbatim}
\end{mybody}
\captionof{table}{Commands and errors for setting up 3DDFA\_V2 repository using LLaMA 3.1, Claude 3 and Mistral.}
\label{tab:commands_2}
\end{table*}

\end{document}